\documentclass[twocolumn,amsmath,trackchanges]{aastex702}

\begin{document}
%\title{Poynting Flux Dissipation and Its Impact on Radiation Transport
%in Strongly Magnetized Supercritical Accretion Flows}

% \title{Effects of Magnetic Flux and Black Hole Spin on Electromagnetic Energy Dissipation and Radiation Transport in Supercritical Accretion Flows}

\title{Radiative Efficiency Enhancement by Electromagnetic Energy Dissipation in Strongly Magnetized Supercritical Accretion Flows around Kerr Black Holes}

------------------------------
\author[orcid=0000-0003-0114-5378,sname='Takahashi']{Hiroyuki R. Takahashi}          
%\altaffiliation{Komazawa University}
\affiliation{Department of Natural Sciences,
Faculty of Arts and Sciences,
Komazawa University,
Setagaya 154-8525, Japan}
\email[show]{takhshhr@komazawa-u.ac.jp}

\author[orcid=0000-0002-2309-3639,sname='Ohsuga']{Ken Ohsuga}          
%\altaffiliation{University of Tsukuba}
\affiliation{Center for Computational Sciences, University of Tsukuba, Tsukuba 305-8577, Japan}
\email{}

\author[orcid=0000-0003-2175-9828,sname='Kobayashi']{Ryohei Kobayashi}          
%\altaffiliation{Institute of Science Tokyo}
\affiliation{Supercomputing Research Center, Institute of Integrated Research, Institute of Science Tokyo, Yokohama, Kanagawa 226–8501, Japan}
\email{}

\author[0000-0002-0700-2223]{Akihiro Inoue}
\affiliation{Department of Earth Science and Astronomy, The University of Tokyo, Meguro, Tokyo 153-8902, Japan}
\email{}

\author[orcid=0000-0003-3640-1749,sname='Asahina']{Yuta Asahina}          
%\altaffiliation{University of Tsukuba}
\affiliation{Center for Computational Sciences, University of Tsukuba, Tsukuba 305-8577, Japan}
\email{}

\author[orcid=0000-0001-7959-6975,sname='Nukada']{Akira Nukada}          
%\altaffiliation{University of Tsukuba}
\affiliation{Center for Computational Sciences, University of Tsukuba, Tsukuba 305-8577, Japan}
\email{}
              
\author[orcid=0000-0001-8730-2228,sname='Boku']{Taisuke Boku}          
%\altaffiliation{Advanced HPC-AI Research and Development Support Center}
\affiliation{Advanced HPC-AI Research and Development Support Center, RIST, Kobe 650-0047, Japan}
%\affiliation{Advanced HPC-AI Research and Development Support Center, Kobe HAIRDESC, Japan}
\email{}
%------------------------------

%\collaboration{all}{The Terra Mater collaboration}

\begin{abstract}
% In black hole supercritical accretion flows, the gravitational energy released by accretion is the primary energy source. However, for rapidly spinning black holes threaded by strong magnetic flux, electromagnetic energy extracted by the Blandford--Znajek mechanism may also make a substantial contribution to energy transport and radiative efficiency. Nevertheless, it remains unclear where the rotational energy of the black hole extracted by the Blandford--Znajek mechanism is dissipated in supercritical accretion flows, and how much it contributes to the radiative luminosity.

We investigate how black hole spin and the amount of magnetic flux affect electromagnetic energy dissipation and radiation transport in supercritical accretion flows.
For this purpose, we perform general relativistic radiation magnetohydrodynamic simulations of MAD and SANE accretion flows with different black hole spins.
In the high-spin MAD model, we find that a fraction of the electromagnetic energy extracted by the Blandford--Znajek mechanism is dissipated near the disk surface in the vicinity of the black hole.
This dissipation significantly contributes to the generation of radiative energy and enhances the luminosity.
The time-averaged radiative efficiency reaches $\eta_{\rm rad}=0.60$, which is much larger than $0.088$ for the non-spinning black hole case and $0.21$ for the weak-magnetic-flux case, corresponding to the SANE state.
As a result, the effective trapping radius, defined as the radius at which the outward radiative luminosity becomes equal to the inward radiative luminosity, is $r_{\rm trap}=2.5r_{\rm g}$, comparable to the ISCO radius.
This value is significantly smaller than $12r_{\rm g}$ for the non-spinning case and $8.5r_{\rm g}$ for the SANE state.
These results suggest that the amount of magnetic flux accumulated on the black hole can affect the radiative properties of supercritical accretion flows and should therefore be considered, in addition to black hole mass, spin, and mass accretion rate, when interpreting observed luminosities and spectra.
\end{abstract}

%\keywords{\uat{Galaxies}{573} --- \uat{Cosmology}{343} --- \uat{High Energy astrophysics}{739} --- \uat{Interstellar medium}{847} --- \uat{Stellar astronomy}{1583} --- \uat{Solar physics}{1476}}

\section{Introduction}
%--------------------------------------------%
% 1
%--------------------------------------------%
Black hole accretion disks are among the most luminous astrophysical phenomena in the Universe and are powered primarily by the gravitational energy released through gas accretion. In particular, supercritical accretion, in which the accretion rate exceeds the Eddington limit, is important for understanding ultraluminous X-ray sources and ultraluminous X-ray pulsars \citep{KaaretUltraluminousRaySources2016,KingUltraluminousRaySources2023}, the rapid growth of supermassive black holes \citep{BegelmanFormationSupermassiveBlack2006,VolonteriOriginsMassiveBlack2021}, and tidal disruption events \citep{GezariTidalDisruptionEvents2021}. In such accretion flows, a large amount of gravitational energy is released. To determine the observed luminosities and spectra, it is therefore crucial to understand where this energy is dissipated, how it is converted into radiation, and how the radiation is transported to large distances.

%--------------------------------------------%
% 2
%--------------------------------------------%
In classical accretion disk models constructed on the basis of one-dimensional analyses, the released energy is assumed to be dissipated mainly inside the disk 
\citep{ShakuraBlackHolesBinary1973,NarayanAdvectionDominatedAccretion1994,AbramowiczSlimAccretionDisks1988}. 
However, high-energy X-ray components that cannot be explained solely by thermal emission from accretion disks are widely observed in black hole X-ray binaries 
\citep{McClintockBlackHoleBinaries2006,DoneModellingBehaviourAccretion2007} 
and Active Galactic Nuclei (AGN) 
\citep{HaardtATwoPhaseModel1991,HaardtXRaySpectraFrom1993,GeorgeXRayReflectionFrom1991,FabianXrayReflection2010}. 
These components include Comptonized emission produced by hot electrons in a corona and reflection components reprocessed at the disk surface. This suggests that a fraction of the accretion energy may be dissipated not only inside the disk, but also near the disk surface or in a hot region above the disk.

%--------------------------------------------%
% 3
%--------------------------------------------%
Theoretically, it has been proposed that magnetic energy generated inside the disk can be transported upward as Poynting flux by magnetic buoyancy and then dissipated near the disk surface or in the coronal region through processes such as magnetic reconnection 
\citep{GaleevStructuredCoronaeOfAccretion1979,MerloniCoronalOutflowDominatedAccretion2002,UzdenskyVerticalStructureAndCoronal2013}. 
Such vertical energy transport and dissipation near the disk surface have also been demonstrated by radiation magnetohydrodynamic (RMHD) simulations. Local RMHD simulations have shown that magnetic fields and energy are transported to the upper layers of the disk by magnetic buoyancy, and that dissipation is distributed not only near the midplane but also near the disk surface 
\citep{TurnerOnTheVerticalStructure2004,HiroseVerticalStructureGas2006,HiroseTurbulentStressesLocal2009,JiangRadiationMagnetohydrodynamicSimulationsOf2014,JiangGlobalThreeDimensional2014}. 
These results indicate that magnetic fields affect not only angular momentum transport, but also the location of energy dissipation and radiative transfer.

%--------------------------------------------%
% 4
%--------------------------------------------%
In systems considered to be powered by supercritical accretion, such as ultraluminous X-ray sources (ULXs), observations have also suggested the presence of optically thick and relatively cool coronal components \citep{GladstoneTheUltraluminousState2009,VierdayantiXRaySpectralVariability2010,ShidatsuNuSTARAndSwiftObservations2017}. Theoretical models that include magnetic heating in disk winds or in coronal regions above the disk have also been proposed \citep{KawanakaWhatDeterminesTheUnique2021}.
% Therefore, energy transport by magnetic fields and dissipation near the disk surface are also important issues for understanding the radiative properties of supercritical accretion flows.
On the other hand, to understand the energy budget of supercritical accretion flows, it is necessary to consider not only the local vertical structure of the disk, but also the global magnetic field structure, outflows, and energy transport associated with black hole spin.
When a sufficiently large amount of poloidal magnetic flux accumulates near the black hole, the accretion flow can enter a magnetically arrested disk (MAD) state \citep{BisnovatyiKoganTheAccretionOfMatter1974,NarayanMagneticallyArrestedDiskAn2003}.
In such strongly magnetized accretion flows around rapidly spinning black holes,
the rotational energy of the black hole can be extracted as Poynting flux by the Blandford--Znajek mechanism 
\citep{BlandfordElectromagneticExtractionEnergy1977} and supplied to the accretion flow. 
Such electromagnetic energy transport may affect the disk structure and radiative properties. 
Indeed, \citet{McKinneyEfficiencySuperEddington2015} showed, using global general relativistic radiation magnetohydrodynamic (GR-RMHD) simulations of supercritical accretion in the MAD state, that radiative efficiencies higher than those predicted by conventional theoretical models can be achieved. They discussed several possible reasons for this high efficiency, including the enhanced photon escape caused by disk compression due to magnetic-flux accumulation and the transport of radiative energy by magnetized outflows.
In addition, \citet{TakahashiFORMATIONOVERHEATEDREGIONS2016} discussed, using GR-RMHD simulations, that the dissipation of electromagnetic energy supplied by the Blandford--Znajek mechanism may contribute to the heating of hot, low-density regions formed near the black hole, in addition to the dissipation of magnetic energy amplified by the magnetorotational instability. However, it remains unclear how energy is transported and dissipated by Poynting flux, and how much it contributes to the radiative luminosity and radiative efficiency. These processes may also affect the trapping radius, which characterizes supercritical accretion \citep{BegelmanCanASphericallyAccreting1979,OhsugaDoesSlimDisk2002,OhsugaSpectralEnergyDistribution2003}.

%--------------------------------------------%
% 5
%--------------------------------------------%
In this study, we therefore focus primarily on supercritical accretion flows in the MAD state with strong magnetic flux and high black hole spin. We analyze the transport and dissipation of electromagnetic energy and investigate how the Poynting flux extracted by the Blandford--Znajek mechanism contributes to the energy budget and radiative luminosity in the vicinity of the accretion flow. Furthermore, to disentangle the effects of magnetic flux and black hole spin, we compare this model with a low-spin MAD model and a high-spin SANE (standard and normal evolution, \citealt{NarayanGRMHDSimulationsMagnetized2012}) model. Through these comparisons, we clarify how the magnetic flux and black hole spin affect the energy dissipation structure, radiation transport, and radiative luminosity.

%--------------------------------------------%
% 6
%--------------------------------------------%
This paper is organized as follows.
In Section~\ref{method}, we describe the numerical setup.
In Section~\ref{result}, we analyze the dissipation of Poynting flux and the associated radiation transport.
In Section~\ref{discussion}, we discuss the formation of the trapping radius, its relation to outflows, and possible observational implications.
Finally, we summarize our findings in Section~\ref{summary}.

%-----------------------------------------------------%
% Table 1
%-----------------------------------------------------
\begin{deluxetable}{ccccccccccc}
\tablecaption{Model parameters and time-averaged properties of the three models.
\label{tab:t-average}}
\tablehead{
   Model & $a_*$& $N_\mathrm{mag}$& averaging interval & $\dot M_\mathrm{BH} (L_\mathrm{Edd})$ & $\phi$ & $r_\mathrm{eq} (r_\mathrm{g})$ & $r_\mathrm{trap} (r_\mathrm{g})$ &
 $\eta_\mathrm{mag}$ & $\eta_\mathrm{rad}$ & $\eta_\mathrm{kin}$ \\
  (1) & (2) & (3) & (4) & (5) & (6) & (7) & (8) & (9) & (10) & (11)
}
\startdata
MAD-0.9 & 0.9 & 1 & 130,000-300,000 & 94 & 77 & 50 & 2.5 & 1.3& 0.60 & 0.13\\
MAD-0 & 0 & 1 & 130,000-195,000& 87 & 72 & 32 & 12 & 0.021 &  0.088 & 0.097\\
SANE-0.9 & 0.9 & 2 & 150,000-300,000  & 59 & 31 &  62 & 8.5 & 0.31 & 0.21 & 0.20\\
\enddata
\tablecomments{
From left to right, (1) the table lists the model name, (2) spin parameter, (3) number of magnetic loops in the $\theta$ direction at the initial state, (4) the time interval used for averaging,
(5) the mass accretion rate onto the black hole $\dot M_\mathrm{BH}$ normalized by the Eddington rate,
(6) the MAD parameter $\phi$, 
(7) the equilibrium radius $r_\mathrm{eq}$,
(8) the trapping radius $r_\mathrm{trap}$,
and the efficiencies of (9) magnetic power $\eta_\mathrm{mag}$,
(10) radiative luminosity $\eta_\mathrm{rad}$, 
and (11) kinetic outflow $\eta_\mathrm{kin}$ measured at $r_\mathrm{eq}$.   
}
\end{deluxetable}

%********************************************%
% section 2
%********************************************%
\section{Numerical method and initial setting}\label{method}
%-----------------------------------------------------
%-----------------------------------------------------
Hereafter, we set the speed of light to $c=1$.
Greek indices, such as $\mu$ and $\nu$, denote spacetime components, whereas Latin indices, such as $i$ and $j$, denote spatial components.
We employ the general relativistic radiation magnetohydrodynamic (GR-RMHD) code UWABAMI \citep{TakahashiFORMATIONOVERHEATEDREGIONS2016}.
This code has been used in previous GR-RMHD studies and has also been upgraded to support GPU acceleration using CUDA-aware MPI and CUDA Fortran \citep{KobayashiAcceleratingGeneralRelativistic2025,KobayashiGPUAcceleratedGeneral2026}.

%%----------------------------------------------------
%%----------------------------------------------------
We solve the GR-RMHD equations in spherical polar coordinates $(r,\theta,\varphi)$.
The number of numerical grid points is $(384,\ 256,\ 1)$.
Because we assume axisymmetry around the polar axis, the dynamo process cannot be self-sustained \citep{CowlingMagneticFieldSunspots1933}.
To compensate for this limitation and study the global long-term evolution of the accretion flow, we solve the GR-RMHD equations with the phenomenological mean-field dynamo model \citep{SadowskiGlobalSimulationsAxisymmetric2015}.

%%----------------------------------------------------
%%----------------------------------------------------
The radial grid spacing increases exponentially with radius.
The inner and outer radial boundaries are located at the black hole horizon $r_{\mathrm{hor}}$ and at $r_{\mathrm{o}} = 5000
r_{\mathrm{g}}$, respectively.
We adopt a non-uniform polar grid defined by
\begin{eqnarray}
\theta = \pi x_2 + \frac{1 - h(r)}{2}\sin(2\pi x_2),
\end{eqnarray}
where $x_2$ is a uniformly spaced coordinate in the interval $[0,1]$ \citep{McKinneyGeneralRelativisticMagnetohydrodynamic2006}.
The function $h(r)$ is given by
\begin{eqnarray}
h(r) &=& A \tanh\left[\frac{r - r_0}{\Delta r}\right] + B, \\
A &=& \frac{h_{\mathrm{o}} - h_{\mathrm{i}}}
{\tanh\left[(r_{\mathrm{o}} - r_0)/\Delta r\right]
- \tanh\left[(r_{\mathrm{hor}} - r_0)/\Delta r\right]}, \\
B &=& \frac{h_{\mathrm{i}}\tanh\left[(r_{\mathrm{o}} - r_0)/\Delta r\right]
- h_{\mathrm{o}}\tanh\left[(r_{\mathrm{hor}} - r_0)/\Delta r\right]}
{\tanh\left[(r_{\mathrm{o}} - r_0)/\Delta r\right]
- \tanh\left[(r_{\mathrm{hor}} - r_0)/\Delta r\right]}.
\end{eqnarray}
With this prescription, $h(r)$ approaches $h_{\mathrm{i}}$ near the horizon, $r = r_{\mathrm{hor}}$,
and $h_{\mathrm{o}}$ at the outer boundary, $r = r_{\mathrm{o}}$.
We set $r_0 = \Delta r = 5\ r_{\mathrm{g}}$, $h_{\mathrm{i}} = 0.6$, and $h_{\mathrm{o}} = 0.1$.
We impose outflow boundary conditions at $r = r_{\mathrm{hor}}$ and $r = r_{\mathrm{o}}$,
and reflective boundary conditions at $\theta = 0$ and $\theta = \pi$.

%%----------------------------------------------------
%%----------------------------------------------------
We initialize the simulation with a Fishbone--Moncrief hydrostatic torus \citep{FishboneRelativisticFluidDisks1976},
which is originally formulated for purely hydrodynamic flows.
To apply this solution to RMHD, we replace the gas pressure with the sum of the gas and radiation pressures,
assuming local thermodynamic equilibrium between the gas and radiation.
The inner edge of the torus is located at $r = r_{\mathrm{edge}} = 200r_{\mathrm{g}}$,
and the pressure maximum is located at $r = r_{\mathrm{peak}}=300r_{\mathrm{g}}$.
The maximum density of the torus is set to
$\rho_0 = 10^{-2}\ \mathrm{g\ cm^{-3}}$.
For numerical stability, we impose a density floor, $\rho_{\mathrm{floor}}=10^{-6}\rho_0 (r/r_{\mathrm{g}})^{-2.5}$, and a gas pressure floor, $p_{\mathrm{gas,floor}}=10^{-8} \rho_0 (r/r_{\mathrm{g}})^{-3.5}$.
We also impose an additional density floor to keep the magnetization parameter $\sigma$ below $\sigma_{\mathrm{max}}=100$.

%%----------------------------------------------------
%%----------------------------------------------------
The inner edge of the torus and the radius of the pressure maximum are approximately an order of magnitude larger than those adopted in our previous simulations and in other studies \citep[e.g.,][]{McKinneyEfficiencySuperEddington2015,SadowskiGlobalSimulationsAxisymmetric2015,FragileLongTimeScale2025,ZhangRadiationGRMHDModels2025}.
This choice is made to minimize the influence of the initial torus and is also motivated by recent studies emphasizing the importance of following accretion flows over sufficiently large spatial scales \citep{KitakiSystematicTwoDimensionalRadiation2018,KitakiOriginsImpactOutflow2021}.
As we show later, the equilibrium radius at which a steady accretion disk forms is smaller than $r_{\mathrm{edge}}$.
Consequently, the equilibrium disk is spatially separated from the initial torus, and its structure is expected to depend only weakly on the detailed properties of the initial torus.

%%----------------------------------------------------
%%----------------------------------------------------
We impose an initial magnetic field on the initial torus.
Following \citet{NathanailPlasmoidFormationGlobal2020}, we prescribe the toroidal component of the vector potential, $A_\varphi$, as
\begin{equation}
A_\varphi=
\mathrm{H}\left(\frac{\rho}{\rho_0}-0.2\right)
\cos \left[(N_{\mathrm{mag}}-1)\theta\right]
\sin \left[\frac{2\pi (r - 1.1r_{\mathrm{edge}})}{\lambda_{\mathrm{mag}}}\right],
\label{eq:Avec}
\end{equation}
where $\rho$ is the mass density and $\mathrm{H}(x)$ is the Heaviside function.
This prescription confines the magnetic field to the high-density region of the torus.
We set the wavelength of the magnetic-field modulation to $\lambda_{\mathrm{mag}} = 100r_{\mathrm{g}}$.
The initial magnetic-field strength is normalized such that the minimum plasma beta, defined as the ratio of the total gas-plus-radiation pressure to the magnetic pressure, is $100$.

%%----------------------------------------------------
%%----------------------------------------------------
We include electron scattering and free--free absorption as sources of opacity.
Compton scattering is included as an energy and momentum exchange process between gas and radiation 

\citep{SadowskiGlobalSimulationsAxisymmetric2015}.
The electron-scattering opacity is assumed to be constant, $\kappa_{\mathrm{es}} = 0.4\ \mathrm{cm^2\ g^{-1}}$.
The free--free absorption opacity is given by
$
\kappa_{\mathrm{ff}} = 6.4 \times 10^{22}\rho T_{\mathrm{gas}}^{-7/2}\ \mathrm{cm^2\ g^{-1}},
$
where $T_{\mathrm{gas}}$ is the gas temperature.
We assume an ideal equation of state with an adiabatic index of $\Gamma=5/3$.
For simplicity, the gas and electron temperatures are assumed to be equal.

%%----------------------------------------------------
%%----------------------------------------------------
We set the black hole mass to $M_{\mathrm{BH}} = 10M_\odot$ throughout this study.
We perform three simulations with different values of the black hole spin parameter $a_*$ and the magnetic-field configuration parameter $N_{\mathrm{mag}}$, defined in Equation~(\ref{eq:Avec}).
The model names and their time-averaged properties are summarized in Table~\ref{tab:t-average}.
For the MAD-0.9 and SANE-0.9 models, we adopt a black hole spin of $a_* = 0.9$, whereas we adopt $a_* = 0$ for the MAD-0 model.
The difference between the MAD and SANE models arises from the choice of $N_{\mathrm{mag}}$.
For the MAD models, we adopt $N_{\mathrm{mag}} = 1$, which produces a magnetic field configuration that threads the equatorial plane.
As a result, a large amount of magnetic flux is efficiently accumulated onto the black hole.
For the SANE model, we adopt $N_{\mathrm{mag}} = 2$, for which the magnetic loops are divided at the equatorial plane, forming symmetric magnetic-loop structures in the northern and southern hemispheres.
This configuration reduces the net poloidal magnetic flux supplied to the black hole compared with the MAD models, and therefore leads to a smaller accumulated magnetic flux on the black hole \citep[see also][]{NathanailMagneticReconnectionPlasmoid2022}.

%%----------------------------------------------------
%% 図1
%%----------------------------------------------------
\begin{figure}
 \begin{center}
    \includegraphics[width=7cm]{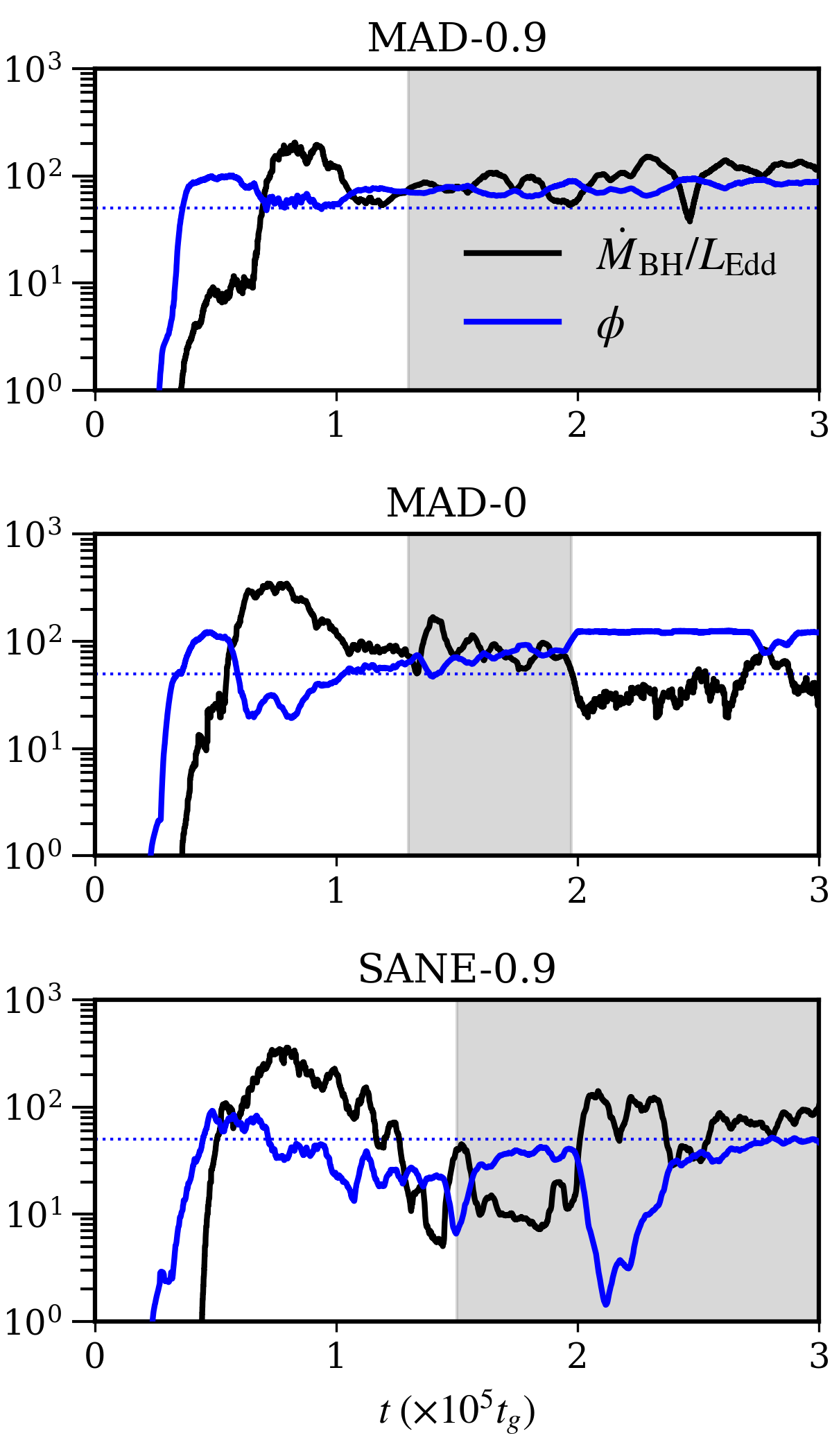} 
        \caption{
        Time evolution of the mass accretion rate $\dot{M}_\mathrm{BH}$ and the MAD parameter $\phi$. The panels show the results for MAD-0.9, MAD-0, and SANE-0.9 from top to bottom, respectively. The horizontal blue dotted line indicates $\phi=50$, which is used as a reference value for the MAD state. 
        The gray shaded regions indicate the time intervals over which time-averaged quantities are evaluated.
        }
    \label{fig:t-mdot}
 \end{center}
\end{figure}

%%----------------------------------------------------
%%----------------------------------------------------

%********************************************%
% section 3.1
%********************************************%
\section{Results}\label{result}
\subsection{Quasi-steady state and time window for averaging}\label{overview}
%--------------------------------------------%
% 1
%--------------------------------------------%
We define the mass inflow rate and mass outflow rate as
\begin{equation}
\dot M_\mathrm{in} =-2\pi \int_0^\pi \rho u^r \mathrm{H}(-u^r)\sqrt{-g} d\theta,
\end{equation}
and
\begin{equation}
\dot M_\mathrm{out} =2\pi \int_0^\pi \rho u^r \mathrm{H}(u^r) \sqrt{-g}d\theta.
\end{equation}
Here, $u^\mu$ and $g$ denote the four-velocity and the determinant of the metric $g_{\mu\nu}$, respectively.
The mass accretion rate onto the black hole is then defined as $\dot M_\mathrm{BH}= \dot M_\mathrm{in}(r_\mathrm{hor}) -
\dot M_\mathrm{out}(r_\mathrm{hor})$.
%-----------------------------------
% Figure 2
%-----------------------------------
\begin{figure*}
 \begin{center}
    \includegraphics[width=19cm]{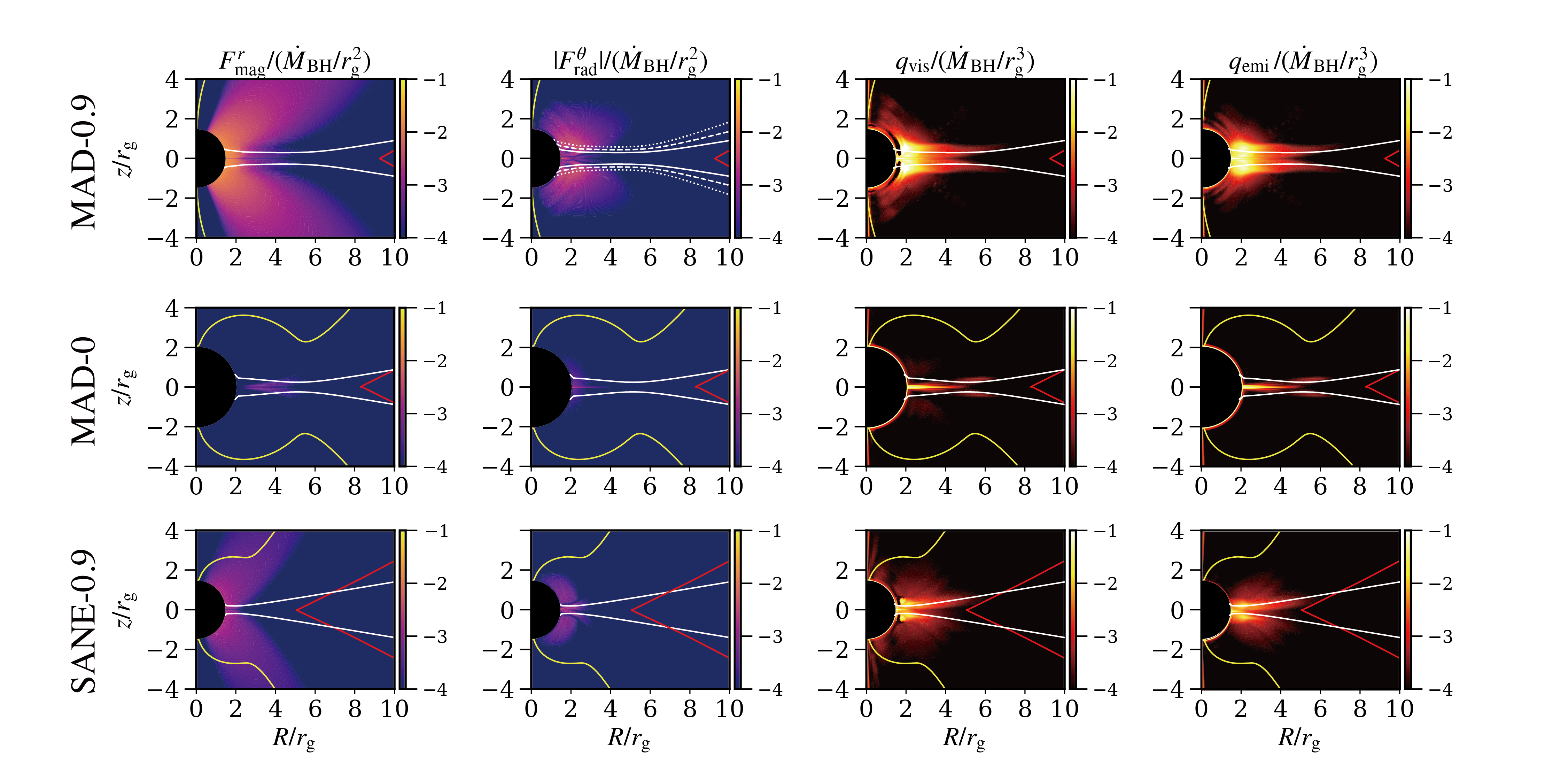} 
        \caption{
From left to right, the color contours show $F_\mathrm{mag}^{r}$, $\left|F_\mathrm{rad}^{\theta}\right|$, $q_\mathrm{vis}$, and $q_\mathrm{emi}$ in the $R$--$z$ plane.
The fluxes are normalized by $\dot M_\mathrm{BH}/r_\mathrm{g}^2$, and the heating and emission rates by $\dot M_\mathrm{BH}/r_\mathrm{g}^3$.
The top, middle, and bottom rows correspond to the MAD-0.9, MAD-0, and SANE-0.9 models, respectively.
The yellow and red curves indicate the surfaces where $\tau_\mathrm{tot}=1$ and $\tau_\mathrm{eff}=1$, respectively.
The white solid curves show the disk surface, $\theta = \pi/2 \pm \theta_H$.
In the second column of the top row, the white dashed and dotted lines indicate the surfaces at $|\theta-\pi/2|=1.5\theta_H$ and $|\theta-\pi/2|=2\theta_H$, respectively.
        }
    \label{fig:pl2d-fq}
 \end{center}
\end{figure*}

%--------------------------------------------%
% 2
%--------------------------------------------%
Figure \ref{fig:t-mdot} shows the time evolution of the mass accretion rate onto the black hole, $\dot M_\mathrm{BH}$, and the MAD parameter $\phi$ for the three models, MAD-0.9, MAD-0, and SANE-0.9. Here, $t_\mathrm{g}$ is the light-crossing time of the gravitational radius $r_\mathrm{g}$. The MAD parameter is defined as
\begin{equation}
\phi = \frac{\left.  2\pi \int_0^{\pi} \frac{1}{2}|B^r|\sqrt{-g}d\theta \right|_{r_\mathrm{hor}}}{\sqrt{\dot M_\mathrm{BH}r_\mathrm{g}^2}},
\end{equation}
where $B^r$ is the radial component of the magnetic three-vector. In all models, after the initial transient phase, both $\dot M_\mathrm{BH}$ and $\phi$ reach a quasi-steady state.

%--------------------------------------------%
% 3
%--------------------------------------------%
In the MAD models, $\phi$ reaches values of $\gtrsim 50$, exceeding the commonly used criterion for the MAD state \citep{TchekhovskoyEfficientGenerationJets2011}.
This indicates substantial magnetic-flux accumulation near the black hole.
In contrast, in the SANE-0.9 model, $\phi$ temporarily increases as $\dot{M}_\mathrm{BH}$ decreases, but it mostly satisfies $\phi<50$ throughout the simulation.

%--------------------------------------------%
% 4
%--------------------------------------------%
All models are evolved until $t=300{,}000t_\mathrm{g}$. In the MAD-0 model, however, after $t=195{,}000t_\mathrm{g}$, the magnetization parameter near the black hole reached the upper limit $\sigma_\mathrm{max}=100$, which was imposed for numerical stability. This causes artificial time variability in the mass accretion rate. We therefore exclude the data after this time from the analysis.
In the following analysis, we use time-averaged data over
$t=130{,}000\text{--}300{,}000\,t_\mathrm{g}$ for MAD-0.9,
$t=130{,}000\text{--}195{,}000\,t_\mathrm{g}$ for MAD-0, and
$t=150{,}000\text{--}300{,}000\,t_\mathrm{g}$ for SANE-0.9.
For the MAD-0 model, the time interval used for averaging is shorter than that for the other models. This time interval approximately corresponds to the viscous time at $r = 40r_\mathrm{g}$, assuming a viscosity parameter of $\alpha = 0.1$ and a disk aspect ratio of $H/r = 0.3$. This radius is comparable to the equilibrium radius $r_\mathrm{eq}$ shown later. Therefore, the system is expected to reach equilibrium inside this radius, and the shorter averaging interval for the MAD-0 model does not significantly affect the discussion in this study.

%--------------------------------------------%
% 5
%--------------------------------------------%
Table~\ref{tab:t-average} summarizes the physical quantities averaged over these time intervals. In all models, the mass accretion rate onto the black hole satisfies $\dot M_\mathrm{BH} \gg L_\mathrm{Edd}$, confirming that the accretion flows studied here are in the supercritical accretion regime.

%--------------------------------------------%
% 6
%--------------------------------------------%
We regard the accretion flow as quasi-steady where the net accretion rate, $\dot M_\mathrm{in}-\dot M_\mathrm{out}$, is nearly independent of radius. We quantify this condition by
\begin{equation}
\xi \equiv \frac{d\ln[(\dot M_\mathrm{in} - \dot M_\mathrm{out})/L_\mathrm{Edd}]}{d\ln(r/r_\mathrm{g})}
\end{equation}
for $r>3r_\mathrm{g}$, and define the equilibrium radius $r_\mathrm{eq}$ as the smallest radius where $\xi=0.3$.
We confirm that the value of $r_\mathrm{eq}$ does not strongly depend on the choice of $\xi$. The values of $r_\mathrm{eq}$ are also listed in Table~\ref{tab:t-average}.

%--------------------------------------------%
% 7
%--------------------------------------------%
We define the inward and outward radiative luminosities, $L_\mathrm{rad}^\mathrm{in}$ and $L_\mathrm{rad}^\mathrm{out}$, respectively, as
\begin{equation}
L_\mathrm{rad}^\mathrm{in} = 2\pi \int_0^\pi R_t^r \mathrm{H}( R_t^r) \sqrt{-g} d\theta,
\end{equation}
and
\begin{equation}
L_\mathrm{rad}^\mathrm{out} =-2\pi \int_0^\pi R_t^r \mathrm{H}(-R_t^r)
\sqrt{-g} d\theta,
\end{equation}
where $R_\mu^\nu$ is the radiation energy-momentum tensor. In this study, we define the trapping radius $r_\mathrm{trap}$ as the smallest radius at which $L_\mathrm{rad}^\mathrm{in}=L_\mathrm{rad}^\mathrm{out}$ is satisfied 
(e.g., \citealt{SadowskiGlobalSimulationsAxisymmetric2015}). The values of $r_\mathrm{trap}$ are listed in Table~\ref{tab:t-average}. We emphasize that $r_\mathrm{eq} > r_\mathrm{trap}$ holds in the radial range considered in this study, and that we evaluate the structure of the accretion flows in regions that reach a steady state.

%--------------------------------------------%
% 8
%--------------------------------------------%
Table~\ref{tab:t-average} also lists the energy conversion efficiencies $\eta_\mathrm{mag}$, $\eta_\mathrm{rad}$, and $\eta_\mathrm{kin}$.
These efficiencies are normalized by $\dot M_\mathrm{BH}$ and are calculated from the magnetic power, radiative luminosity, and kinetic energy power measured at $r=r_\mathrm{eq}$, respectively.

%--------------------------------------------%
% 9
%--------------------------------------------%
In the following sections, we discuss the time-averaged structures over the time intervals listed in Table~\ref{tab:t-average}. The time variability is discussed in Section~\ref{sec:t-r}.

%********************************************%
% section 3.2
%********************************************%
\subsection{Disk heating by dissipating the Poynting flux}\label{disp}
%--------------------------------------------%
% 1
%--------------------------------------------%
Figure~\ref{fig:pl2d-fq} shows contour plots in the $R$--$z$ plane,
where $R=r\sin\theta$ and $z=r\cos\theta$.
The rows correspond to MAD-0.9, MAD-0, and SANE-0.9, respectively.
The columns show the radial Poynting flux $F_\mathrm{mag}^{r}$,
the absolute value of the polar radiative flux $\left|F_\mathrm{rad}^{\theta}\right|$,
the effective viscous heating rate per unit volume $q_\mathrm{vis}$,
and the effective emission rate per unit volume $q_\mathrm{emi}$.
These quantities are normalized by $\dot M_\mathrm{BH}/r_\mathrm{g}^2$ for the fluxes
and by $\dot M_\mathrm{BH}/r_\mathrm{g}^3$ for the heating and emission rates per unit volume.
Here, $q_\mathrm{vis}$ represents the effective heating rate associated with viscous and magnetic dissipation, while $q_\mathrm{emi}$ represents the effective rate of energy transfer from gas to radiation through radiative processes.
The detailed definitions of these quantities are given in Appendix A.
The yellow and red curves indicate the surfaces on which the total optical depth
$\tau_\mathrm{tot}$ and the effective optical depth $\tau_\mathrm{eff}$,
measured from the polar axis, are equal to unity, respectively.
The white curves denote the surfaces corresponding to the density scale height,
defined as
\begin{equation}
\theta_H=\frac{1}{2}
\sqrt{
\frac{\int_0^{\pi}\rho (\theta-\pi/2)^2\sqrt{g_{\theta \theta}}d\theta}
{\int_0^{\pi}\rho \sqrt{g_{\theta \theta}}d\theta}
}.
\end{equation}
In the following analysis, we use $\theta=\pi/2\pm\theta_H$ as the disk-surface locations.
%%In the following, we set $\theta'=\pi/2\pm\theta_{H}$.

%--------------------------------------------%
% 2
%--------------------------------------------%
In the MAD-0.9 model, a strong Poynting flux is present above the disk surface. This is consistent with the picture in which the outward electromagnetic energy flux generated by the Blandford--Znajek mechanism is transported mainly in the polar direction 
\citep{BlandfordElectromagneticExtractionEnergy1977}. 
Indeed, we confirm that the electromagnetic energy flux measured at the horizon is in good agreement with the theoretical prediction of Blandford--Znajek \citep{McKinneyMeasurementElectromagneticLuminosity2004}, indicating that the outward Poynting flux originates mainly from the Blandford--Znajek mechanism.
A similar outward Poynting flux at the horizon is also seen in SANE-0.9, although it is weaker than that in MAD-0.9. This is because the value of $\phi$ in SANE-0.9 is smaller than that in MAD-0.9 (Table~\ref{tab:t-average}), and the Blandford--Znajek mechanism therefore operates less efficiently \citep{TchekhovskoyBlackHoleSpin2010}. 
In contrast, no comparable Poynting flux is seen in the MAD-0 model.
%-----------------------------------------------------%
% Figure 3, 模式図
%-----------------------------------------------------%
\begin{figure}
\begin{center}
\includegraphics[width=8cm]{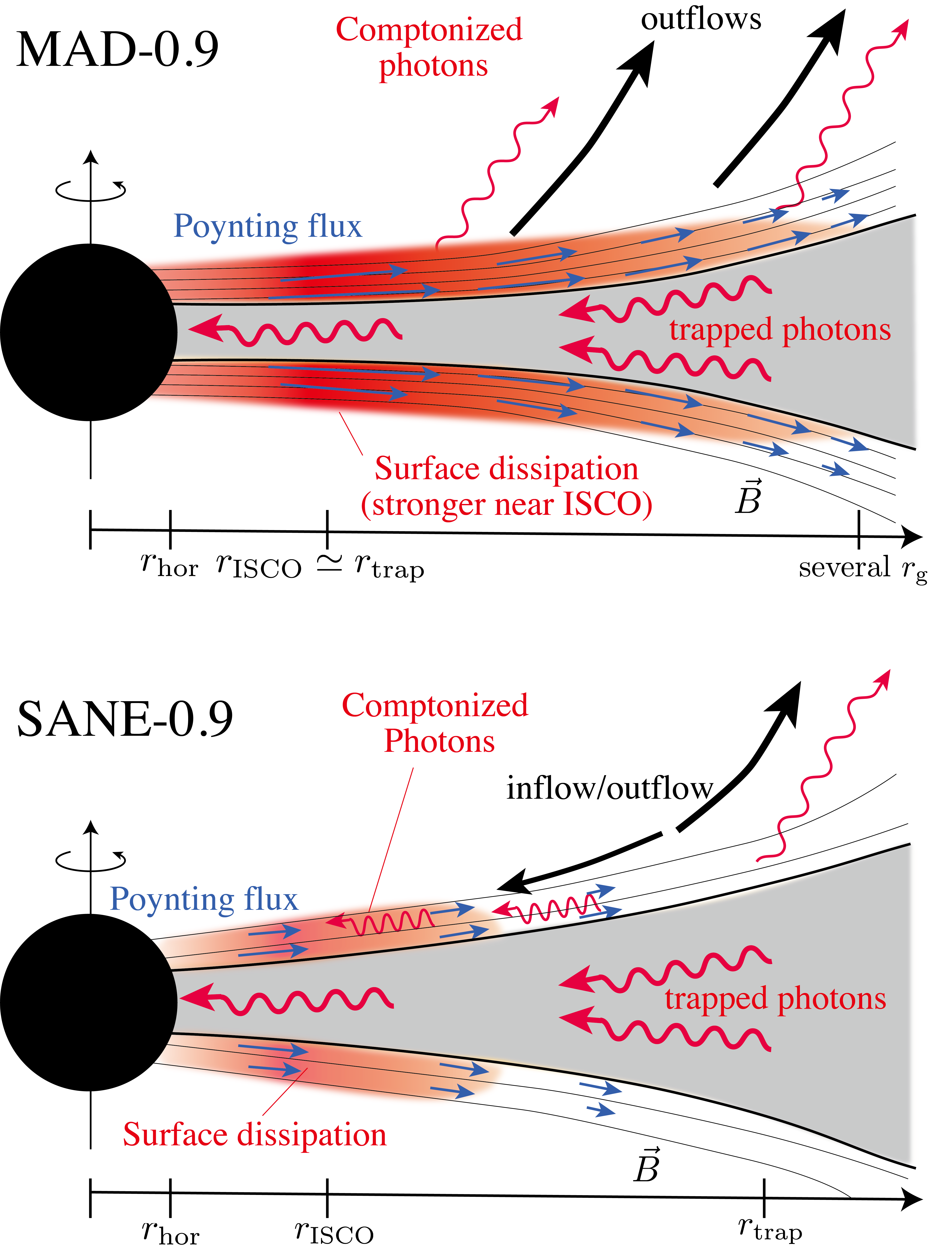}
\caption{
Schematic illustration of energy dissipation and radiation transport in the MAD-0.9 (top) and SANE-0.9 (bottom) models.
Thick black curves denote the disk surface defined by the density scale height.
Blue arrows, thin black curves, red wavy arrows, and thick black arrows represent the Poynting flux, magnetic field lines, radiation transport, and gas outflow/inflow motions, respectively.
Darker red colors qualitatively indicate stronger dissipation above the disk.
In MAD-0.9, strong surface dissipation occurs near the ISCO, and the generated radiation is efficiently transported outward with the outflow.
In SANE-0.9, surface dissipation is weaker, and part of the radiation is advected inward with the accretion flow.
} 
\label{fig:illust}
\end{center}
\end{figure}

%--------------------------------------------%
% 3
%--------------------------------------------%
The third column, which shows $q_\mathrm{vis}$, indicates that strong energy dissipation occurs near the disk surface in the MAD-0.9 model. 
A similar, but weaker, bipolar dissipation structure is also seen on both sides of the disk in the SANE-0.9 model.
In contrast, the MAD-0 model does not show such a bipolar structure, and the dissipation region is confined to the disk interior. These results suggest that, in the MAD-0 model, dissipation driven mainly by turbulence inside the disk is dominant, whereas in the MAD-0.9 model, dissipation of electromagnetic energy associated with the Blandford--Znajek mechanism occurs near the disk surface.

%--------------------------------------------%
% 4
%--------------------------------------------%
The fourth column, which shows $q_\mathrm{emi}$, exhibits a spatial distribution similar to that of $q_\mathrm{vis}$. This indicates that the dissipated energy is promptly converted into radiative energy. Furthermore, the polar radiative flux $F^{\theta}$ shown in the second column indicates that the radiative energy generated in this way is transported upward above the disk. This trend is particularly prominent in the MAD-0.9 model.

%--------------------------------------------%
% 5
%--------------------------------------------%
Figure~\ref{fig:illust} schematically illustrates the overall picture of energy dissipation and radiation transport suggested by the spatial structures shown in Figure~\ref{fig:pl2d-fq}. The upper and lower panels correspond to MAD-0.9 and SANE-0.9, respectively. The thick black curves indicate the disk surface corresponding to the density scale height. The blue arrows, thin black lines, red wavy arrows, and black arrows represent the Poynting flux, magnetic field lines, radiation transport, and gas motion, respectively. The darker red color indicates stronger dissipation.
As shown in Figure~\ref{fig:pl2d-fq}, in MAD-0.9, the Poynting flux originating from the Blandford--Znajek mechanism is transported through the region near the disk surface, where strong energy dissipation occurs. Similar energy dissipation is also seen in SANE-0.9, but the dissipation rate is smaller than that in MAD-0.9.

\begin{figure}
 \begin{center}
    \includegraphics[width=8cm]{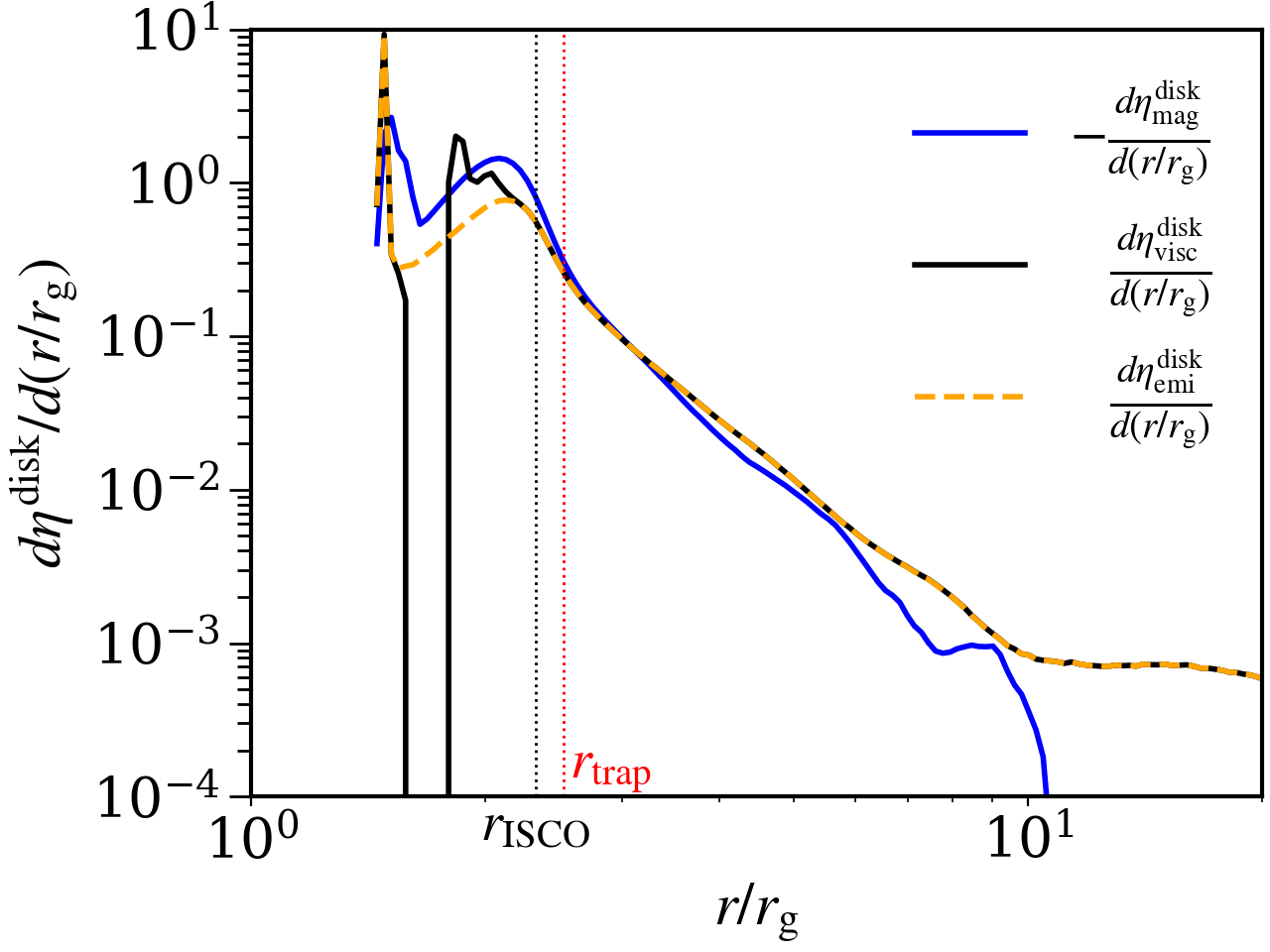}
        \caption{
Radial profiles of the disk-integrated magnetic power decrease rate $-dL_\mathrm{mag}^\mathrm{disk}/dr$ (blue curve), heating rate $dQ_\mathrm{vis}^\mathrm{disk}/dr$ (black curve), and emission rate $dQ_\mathrm{emi}^\mathrm{disk}/dr$ (orange dashed curve) for MAD-0.9.
All quantities are integrated over the disk region and normalized by $\dot M_\mathrm{BH}/r_\mathrm{g}$.
Vertical dotted and dashed lines indicate the ISCO radius and the trapping radius $r_\mathrm{trap}$, respectively.
 }
    \label{fig:r-dLdr}
 \end{center}
\end{figure}

%--------------------------------------------%
% 6
%--------------------------------------------%
In MAD-0.9, the dissipation of electromagnetic energy and the associated radiation transport are most prominent. We therefore focus on this model and quantitatively examine the radial decrease in magnetic power inside the disk and its relation to the associated heating and emission rates.
Using the density scale height $\theta_{H}$, we define the magnetic power inside the disk, $L_\mathrm{mag}^\mathrm{disk}$, as
\begin{equation}
L_\mathrm{mag}^\mathrm{disk} =
2\pi \int_{\pi/2-\theta_H}^{\pi/2+\theta_H}
F_\mathrm{mag}^r \sqrt{-g} d\theta
\end{equation}
and show the radial decrease rate of the magnetic power,
$-dL_\mathrm{mag}^\mathrm{disk}/dr$, by the blue curve in Figure~\ref{fig:r-dLdr}.
This quantity is normalized by $\dot M_\mathrm{BH}/r_\mathrm{g}$.
The vertical dotted lines indicate the ISCO (black) and $r_\mathrm{trap}$ (red).

%-----------------------------------------------------%
%Figure 4　dL/dr
%-----------------------------------------------------%

%--------------------------------------------%
% 10
%--------------------------------------------%
The magnetic power generated near the black hole is supplied to the disk, decreases strongly around the ISCO, and drops rapidly outside the ISCO. The logarithmic slope of the radial profile of the magnetic-power decrease rate is approximately $-4.0$ at $r=2$--$3\,r_\mathrm{g}$ and approximately $-8.5$ at $r=3$--$6\,r_\mathrm{g}$. 
Such a steep radial decline of the disk-integrated magnetic power suggests that the electromagnetic energy extracted by the Blandford--Znajek mechanism 
is converted into other forms of energy inside the disk,
rather than simply being transported outward without dissipation \citep{TakahashiFORMATIONOVERHEATEDREGIONS2016}.

%-----------------------------------------------------%
%Figure 5　dq/dtheta, dL/dr
%-----------------------------------------------------%
\begin{figure*}
 \begin{center}
\includegraphics[width=15cm]{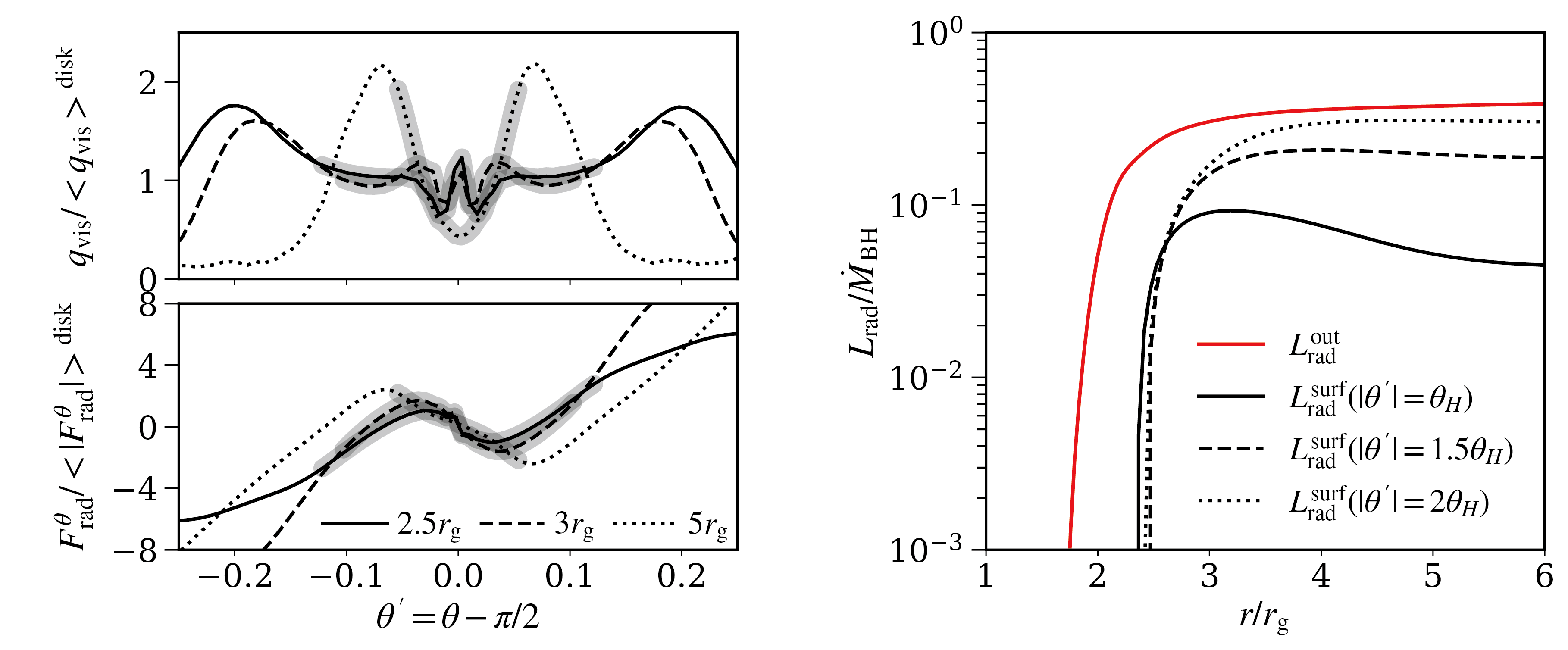}
        \caption{
%Angular and radial profiles of radiative and dissipative quantities in the MAD-0.9 model.
Left: polar-angle dependence of the effective viscous heating rate (top) and the polar radiative flux (bottom) at 
$r = 2.5\,r_\mathrm{g}$ (solid), 
$r = 3\,r_\mathrm{g}$ (dashed), 
and $r = 5\,r_\mathrm{g}$ (dotted).
Each quantity is normalized by its disk-averaged value at the corresponding radius.
Hatched regions indicate the disk interior, defined by $|\theta^\prime| < \theta_H$.
Right: radial profiles of the cumulative luminosity evaluated at the surfaces 
$|\theta^\prime| = \theta_H$ (solid), 
$|\theta^\prime| = 1.5\theta_H$ (dashed), 
and $|\theta^\prime| = 2\theta_H$ (dotted).
Red curve shows the total outward radiative luminosity $L_\mathrm{rad}^\mathrm{out}$.
All luminosities are normalized by the mass accretion rate $\dot M_\mathrm{BH}$.
}
    \label{fig:theta-q_r-L}
 \end{center}
\end{figure*}
%--------------------------------------------%
% 11
%--------------------------------------------%
Next, we define the heating rate and emission rate per unit radius inside the disk,
$dQ_\mathrm{vis}^\mathrm{disk}/dr$ and $dQ_\mathrm{emi}^\mathrm{disk}/dr$, respectively, as
\begin{eqnarray}
\frac{dQ^\mathrm{disk}_\mathrm{vis}}{dr}
&=& 2\pi\int_{\pi/2-\theta_H}^{\pi/2+\theta_H}
q_\mathrm{vis}\sqrt{-g}d\theta,\\
\frac{dQ^\mathrm{disk}_\mathrm{emi}}{dr}
&=&  2\pi\int_{\pi/2-\theta_H}^{\pi/2+\theta_H}
q_\mathrm{emi}\sqrt{-g}d\theta.
\end{eqnarray}
The normalized heating rate,
$\eta_\mathrm{vis}^\mathrm{disk}\equiv
(r_\mathrm{g}/\dot M_\mathrm{BH})
dQ^\mathrm{disk}\mathrm{vis}/dr$ (black),
and the normalized emission rate,
$\eta_\mathrm{emi}^\mathrm{disk}\equiv
(r_\mathrm{g}/\dot M_\mathrm{BH})
dQ^\mathrm{disk}_\mathrm{emi}/dr$ (orange),
are shown in Figure~\ref{fig:r-dLdr}.

% The heating rate $dQ_\mathrm{vis}^\mathrm{disk}/dr$ (black) and the emission rate $dQ_\mathrm{emi}^\mathrm{disk}/dr$ (orange) are shown in Figure~\ref{fig:r-dLdr}. These quantities are normalized by $\dot M_\mathrm{BH}/r_\mathrm{g}$, in the same way as the magnetic power dissipation rate.
This figure shows that, in the range of $r \simeq 2$--$6\,r_\mathrm{g}$, the magnetic-power decrease rate $-dL_\mathrm{mag}^\mathrm{disk}/dr$ agrees well with both $dQ_\mathrm{vis}^\mathrm{disk}/dr$ and $dQ_\mathrm{emi}^\mathrm{disk}/dr$. 
This indicates that the magnetic energy lost inside the disk is converted mainly into gas heating and subsequently into radiative energy.
Thus, near the black hole, the rotational energy extracted from the black hole is supplied to the disk as electromagnetic energy and efficiently converted into radiation.

%--------------------------------------------%
% 12
%--------------------------------------------%
The left panels of Figure~\ref{fig:theta-q_r-L} show the polar-angle dependence of the effective viscous heating rate (upper panel) and the radiative flux in the $\theta$ direction (lower panel) for MAD-0.9. Here, we define $\theta^\prime = \theta-\pi/2$. The solid, dashed, and dotted lines correspond to $r=2.5r_\mathrm{g}$ (near the ISCO), $r=3r_\mathrm{g}$ (where the decrease in magnetic power is significant), and $r=5r_\mathrm{g}$ 
(where the decrease becomes weaker), respectively. The hatched region represents the disk interior, defined by $|\theta^\prime|<\theta_H$.

%--------------------------------------------%
% 13
%--------------------------------------------%
The upper panel shows that the effective heating rate is enhanced near the disk surface rather than being distributed uniformly inside the disk.
This indicates that the magnetic-energy dissipation inferred from the radial decrease in $L_\mathrm{mag}^\mathrm{disk}$ in Figure~\ref{fig:r-dLdr} occurs mainly near the disk surface.

%--------------------------------------------%
% 14
%--------------------------------------------%
The radiative flux in the $\theta$ direction shown in the lower panel increases near the disk surface, corresponding to the dissipation region seen in the upper panel. This suggests that the enhanced heating and radiation near the disk surface are closely related, and that radiative energy is transported in the polar direction. In particular, at $r=2.5$--$3r_\mathrm{g}$, a strong radiative flux is seen upward and downward from the disk surface.

%--------------------------------------------%
% 15
%--------------------------------------------%
The right panel of Figure~\ref{fig:theta-q_r-L} shows the cumulative radiative luminosity passing through the surface $\Sigma$,
\begin{equation}
L_\mathrm{rad}(r)
=-
\int_{\Sigma}
R_t{}^{n}\, d\Sigma_n,
\end{equation}
where $\Sigma$ is the surface region corresponding to $r_\mathrm{hor}\le r' \le r$, and $n$ denotes the direction normal to the surface. The black solid line corresponds to the case in which $\Sigma$ is taken to be the disk surface ($|\theta'|=\theta_H$). The red curve shows $L_\mathrm{rad}^\mathrm{out}$.
The radiative energy generated in the dissipation region is transported above the disk through the disk surface. 
Within the radial range shown in Figure~\ref{fig:theta-q_r-L}, the luminosity passing through the disk surface is $\lesssim 0.1\,\dot M_\mathrm{BH}$, which is only about one quarter of the total outward radiative luminosity $L_\mathrm{rad}^\mathrm{out}$. 
This is because the dissipation region is not localized at the disk surface, but extends above the disk (see Figure~\ref{fig:pl2d-fq}).

%--------------------------------------------%
% 16
%--------------------------------------------%
We therefore check the cumulative radiative luminosities passing through the surfaces located slightly above the disk surface, $|\theta^\prime|=1.5\theta_H$ and $|\theta^\prime|=2\theta_H$. These luminosities are shown by the dashed and dotted lines, respectively, in the right panel of Figure~\ref{fig:theta-q_r-L}. The locations of these surfaces are also shown with the same line styles in the second panel from the left in the top row of Figure~\ref{fig:pl2d-fq}.
As in the case of the disk surface, the cumulative luminosities passing through these surfaces increase rapidly with radius in the region where electromagnetic energy dissipation is dominant. 
The cumulative luminosity also increases with $|\theta^\prime|$ and is nearly saturated at $|\theta^\prime|=2\theta_H$, where it reaches a value comparable to $L_\mathrm{rad}^\mathrm{out}$ at $r\simeq 4r_\mathrm{g}$.
These results indicate that, near the disk surface in the vicinity of the black hole, a major fraction of the radially transported radiative energy is generated in association with electromagnetic energy dissipation. In the MAD-0.9 panel of Figure~\ref{fig:illust}, such electromagnetic energy dissipation near the disk surface, shown by the reddish region, and the associated outward radiation transport, shown by the red wavy arrows, are illustrated schematically.
%-----------------------------------------------------%
% 図6 r方向輻射フラックスのコンター
%-----------------------------------------------------%
\begin{figure}
 \begin{center}
\includegraphics[width=8.5cm]{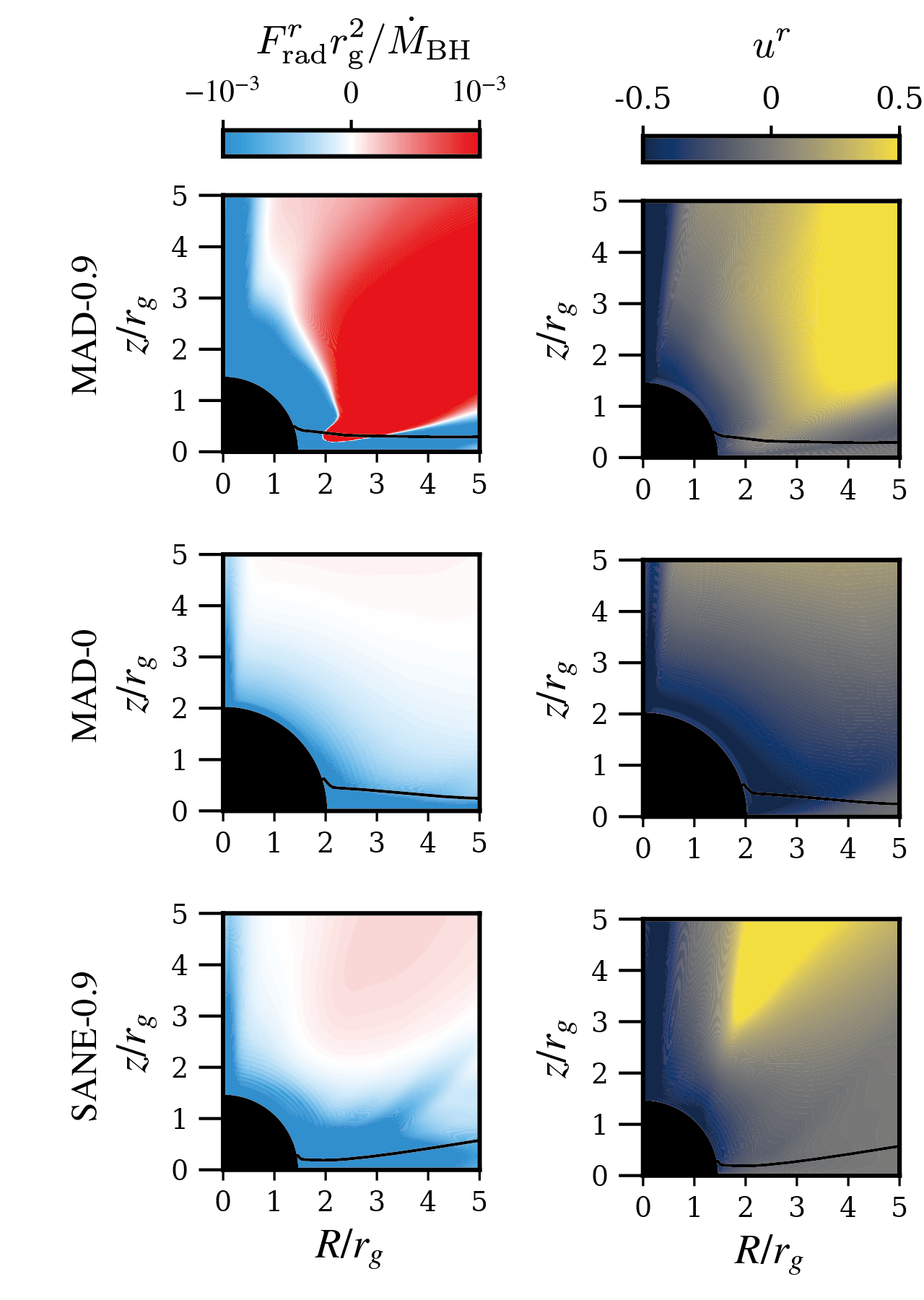}
        \caption{
Left panels show the radial radiative flux $F^r_\mathrm{rad}$ normalized by $\dot M_\mathrm{BH}/r_\mathrm{g}^2$, while right panels show the radial four-velocity $u^r$.
Black curves show the disk surface.         
        }
    \label{fig:frad_cont}
 \end{center}
\end{figure}

%--------------------------------------------%
% 17
%--------------------------------------------%
Here, we note that, as shown in the right panel of Figure~\ref{fig:theta-q_r-L}, the surface luminosity at $|\theta'|=\theta_H$ has a peak at $r \simeq 3 r_\mathrm{g}$ and decreases outside this radius.
Figure~\ref{fig:frad_cont} shows the radiative flux $F_\mathrm{rad}^r$ (left) and the radial component of the four-velocity $u^r$ (right). The black curves indicate the disk surface, $|\theta'|=\theta_H$. From top to bottom, the panels show the results for MAD-0.9, MAD-0, and SANE-0.9, respectively. 
In the MAD-0.9 model, both the radiative flux and the fluid velocity are outward near the disk surface at $R \simeq 2$--$3 r_\mathrm{g}$, where magnetic-energy dissipation is strong.
As indicated by the optical depths shown in Figure~\ref{fig:pl2d-fq}, this region is optically thick. The radiative energy generated by electromagnetic energy dissipation is transported outward together with the outflow and contributes to $L_\mathrm{rad}^\mathrm{out}$ (see the red wavy and black arrows in the top panel of Figure~\ref{fig:illust}). In contrast, at $R \gtrsim 4 r_\mathrm{g}$, the energy supply by magnetic dissipation becomes weaker, and the gas flow is directed inward. 
As a result, radiation advected inward with the fluid becomes dominant, reducing the outward luminosity from the disk surface.

%--------------------------------------------%
% 18
%--------------------------------------------%
In contrast, in MAD-0, neither a clear outward radiative flux nor an outflow is seen near the disk surface. This is consistent with the fact that, in the absence of black hole spin, the supply of electromagnetic energy by the Blandford--Znajek mechanism is negligible and additional radiation production near the disk surface is weak.
In SANE-0.9, a weak outward radiative flux is seen above the disk, but an inward radiative flux and inward fluid velocity remain near the disk surface. Therefore, compared with MAD-0.9, the radiative energy generated by dissipation is less efficiently transported outward.

%-----------------------------------------------------%
%Figure 7 平衡半径とtrapping radius
%-----------------------------------------------------%
\begin{figure}
 \begin{center}
    \includegraphics[width=8cm]{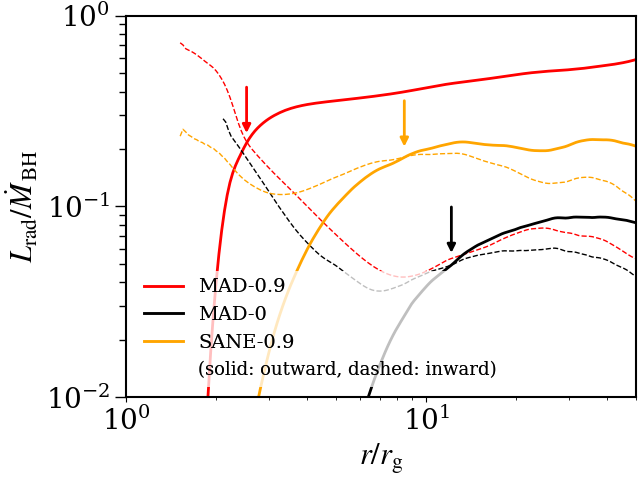} 
         \caption{
Radial profiles of the outward radiative luminosity $L_\mathrm{rad}^\mathrm{out}$ (solid) and inward radiative luminosity $L_\mathrm{rad}^\mathrm{in}$ (dashed), normalized by $\dot{M}_\mathrm{BH}$.
Red, black, and orange curves correspond to the MAD-0.9, MAD-0, and SANE-0.9 models, respectively.
Arrows indicate the trapping radius, defined as the radius where $L_\mathrm{rad}^\mathrm{out} = L_\mathrm{rad}^\mathrm{in}$.
  }
    \label{fig:r-Lrad}
 \end{center}
\end{figure}

%--------------------------------------------%
% 19
%--------------------------------------------%
Figure~\ref{fig:r-Lrad} shows the radial profiles of the time-averaged outward radiative luminosity $L_\mathrm{rad}^\mathrm{out}$ (solid lines) and inward radiative luminosity $L_\mathrm{rad}^\mathrm{in}$ (dashed lines). The red, black, and orange lines correspond to MAD-0.9, MAD-0, and SANE-0.9, respectively. The red solid line in Figure~\ref{fig:r-Lrad} shows the same data as the red solid line in the right panel of Figure~\ref{fig:theta-q_r-L}; it is replotted here to clarify the comparison with the other models.

%--------------------------------------------%
% 20
%--------------------------------------------%
Comparing MAD-0.9 and MAD-0, we find no significant difference in the magnitude or radial distribution of the inward luminosity outside the ISCO. 
This would reflect the fact that the mass accretion rates are comparable in the two models (see Table~\ref{tab:t-average}). In both models, the radiative flux inside the disk is directed inward, indicating that photon trapping is dominant.
On the other hand, in MAD-0.9, the outward luminosity increases significantly near the black hole. As shown above, this enhancement is consistent with the picture in which the Poynting flux supplied by the Blandford--Znajek mechanism undergoes energy conversion near the disk surface and is converted into radiative energy. As a result, in MAD-0.9, the energy supply near the black hole increases and outward radiation transport is enhanced, changing the balance between inward and outward radiation transport. The trapping radius $r_\mathrm{trap}$ therefore becomes smaller and is located close to the ISCO.
Thus, the decrease in $r_\mathrm{trap}$ in MAD-0.9 is not interpreted as a simple weakening of photon trapping. Instead, it reflects the generation of additional radiative energy of Blandford--Znajek origin near the black hole and the resulting enhancement of outward radiation transport.
This enhancement of outward radiation transport is also supported by the radiative efficiencies listed in Table~\ref{tab:t-average}. The radiative efficiency of MAD-0.9, $\eta_\mathrm{rad}=0.60$, is much larger than that of MAD-0, $\eta_\mathrm{rad}=0.088$. The trend that a high radiative efficiency and a small trapping radius are obtained in the high-spin MAD state is qualitatively consistent with the results of \citet{McKinneyEfficiencySuperEddington2015}.

%--------------------------------------------%
% 21
%--------------------------------------------%
%>>>>>>>>>>>>>>>>>>>>>>>>>>>>>>>>>
%>>>>>>>>>>>>>>>>>>>>>>>>>>>>>>>>>
Next, comparing SANE-0.9 with the other models, we find that the inward luminosity in SANE-0.9 is slightly larger than that in the other models. For example, at $r = 10 r_\mathrm{g}$, $L_\mathrm{rad}^\mathrm{in}/\dot{M}_\mathrm{BH}$ in SANE-0.9 is about four times larger than that in the other models. One possible reason for this is that the disk in SANE-0.9 is thicker than those in the other models. At $r = 10 r_\mathrm{g}$, $\theta_H$ is about 1.5 times larger than in the other models, which may enhance the inward radiative luminosity \citep{McKinneyEfficiencySuperEddington2015}. However, as shown in Table~\ref{tab:t-average}, 
$\dot{M}_\mathrm{BH}$ in SANE-0.9 is smaller than those in the other models by about a factor of $2/3$. 
Therefore, our calculations do not necessarily indicate that photon trapping is more efficient in SANE-0.9.
On the other hand, as shown in Figure~\ref{fig:pl2d-fq}, electromagnetic energy dissipation is also seen near the disk surface at $R \simeq 2$--$3r_\mathrm{g}$ in SANE-0.9.
However, the subsequent radiation transport differs from that in MAD-0.9.
Figure~\ref{fig:frad_cont} shows that, in MAD-0.9, both the radiative flux and gas motion are directed outward in this region, whereas in SANE-0.9 an inward radiative flux and inward gas motion remain.
This difference suggests that, in MAD-0.9, a strong Blandford--Znajek-driven outflow transports the radiative energy generated by dissipation outward, while in SANE-0.9, where such a strong outflow is absent, a fraction of the radiative energy is advected toward the black hole, increasing $L_\mathrm{rad}^\mathrm{in}/\dot{M}_\mathrm{BH}$.
This contrast is schematically illustrated in Figure~\ref{fig:illust}: an outflow accompanied by outward radiation transport in MAD-0.9, and an inflow/outflow structure with an inward component in SANE-0.9.

%--------------------------------------------%
% 22
%--------------------------------------------%
Focusing on $L_\mathrm{rad}^\mathrm{out}$ in Figure~\ref{fig:r-Lrad}, we find that SANE-0.9 has a smaller outward radiative luminosity than MAD-0.9, but a larger one than MAD-0. One possible reason is that, even in SANE-0.9, a dissipation region extending above the disk surface is seen at $r \lesssim 3 r_\mathrm{g}$ (Figure~\ref{fig:pl2d-fq}), which may partly contribute to the outward radiation transport. Indeed, Figure~\ref{fig:frad_cont} shows an outward radiative flux around $r \simeq 3r_\mathrm{g}$ in SANE-0.9, whereas no similar feature is seen in MAD-0. Therefore, weak magnetic-energy dissipation in SANE-0.9 may contribute to the increase in the outward radiative luminosity.
Another possible explanation is that the high black hole spin reduces the inner disk radius and increases the released gravitational energy. The ratio of the ISCO radii between SANE-0.9 and MAD-0 is $\sim 2.6$, and this difference may partly contribute to the luminosity difference. However, considering that $L_\mathrm{rad}^\mathrm{in}$ is comparable between MAD-0.9 and MAD-0, which have comparable accretion rates but different ISCO radii, it is difficult to explain the luminosity difference between SANE-0.9 and MAD-0 solely by the difference in the gravitational potential due to spin.

%--------------------------------------------%
% 23
%--------------------------------------------%
These results show that radiation transport is strongly affected by differences in fluid motion, which are determined by black hole spin and magnetic flux.
In particular, outward transport dominates in MAD-0.9, whereas the transport component directed inward is not negligible in SANE-0.9.

%********************************************%
% section 4.1
%********************************************%
\section{discussion}\label{discussion}
\subsection{Temporal correlation between magnetic flux and characteristic radii}
\label{sec:t-r}
%--------------------------------------------%
% 1
%--------------------------------------------%
In Section~\ref{result}, we found that the trapping radius depends on the MAD parameter $\phi$.
Here, we further define the outflow development region and the MAD region, and examine how these regions, together with the trapping region, depend on $\phi$.

% In Section~\ref{result}, we show that, in the MAD-0.9 model, disk heating by magnetic dissipation causes the trapping radius to be located close to the ISCO. In this section, in addition to the trapping radius, we define the radius at which an outflow is launched and the radius of the region where the disk is compressed by magnetic pressure, namely the MAD radius. We then discuss how these radii depend on the MAD parameter $\phi$.

%--------------------------------------------%
% 2
%--------------------------------------------%
In Section~3, we use the radius at which $L_\mathrm{rad}^\mathrm{in} = L_\mathrm{rad}^\mathrm{out}$ is satisfied as a representative value of the trapping radius. 
However, as shown in Figure~\ref{fig:r-Lrad}, especially in the MAD-0 and SANE-0.9 models, both luminosities vary only gradually with radius, making it difficult to identify a unique trapping radius.
In this section, we therefore relax the definition of $r_\mathrm{trap}$ and define the trapping region as the radial range that satisfies
$L_\mathrm{rad}^{\mathrm{out}}/L_\mathrm{rad}^{\mathrm{in}}\in [0.5,\ 1]$.
We denote the representative radius of this region by $r_\mathrm{trap}$.

%--------------------------------------------%
% 3
%--------------------------------------------%
Next, we define the outflow development region.
Similar to $L_\mathrm{rad}^\mathrm{out}$, the mass outflow rate $\dot M_\mathrm{out}$ increases rapidly near the black hole and can exceed $\dot M_\mathrm{BH}$.
Therefore, this substantial outflow is expected to play a non-negligible role in the accretion process.
We define the outflow development region as the radial range where
$\dot M_\mathrm{out}/\dot M_\mathrm{BH}\in[0.5,\ 1]$,
and denote its representative radius by $r_\mathrm{out}$.
This definition identifies the region where the cumulative mass outflow becomes comparable to the mass accretion rate, rather than the location where the outflow is first launched.
Thus, even when outward motion already exists near the black hole, as shown in Figure~\ref{fig:frad_cont}, $r_\mathrm{out}$ can be located farther out if the inner outward-moving region contributes only weakly to $\dot M_\mathrm{out}$.

% Next, we define the outflow development region. 
% Similar to $L_\mathrm{rad}^\mathrm{out}$, the mass outflow rate $\dot M_\mathrm{out}$ also increases with radius. 
% In particular, $\dot M_\mathrm{out}$ increases rapidly near the black hole, and its magnitude exceeds $\dot M_\mathrm{BH}$. Therefore, this massive outflow plays a non-negligible role in the accretion process.
% In this section, we define the outflow development region as the radial range where a massive outflow is formed, satisfying
% $\dot M_\mathrm{out}/\dot M_\mathrm{BH}\in[0.5,\ 1]$.
% We denote the representative scale of this region by $r_\mathrm{out}$. This definition is intended to identify the region where the mass outflow becomes sufficiently large, and does not directly indicate the location where the outflow is first launched. Therefore, even when an outward flow already exists near the black hole, as shown in Figure~\ref{fig:frad_cont}, $r_\mathrm{out}$ can be located farther out if the contribution to the mass outflow rate $\dot M_\mathrm{out}$ is small.

%--------------------------------------------%
% 4
%--------------------------------------------%
Finally, we define the MAD transition region, where a geometrically compressed MAD disk is formed. 
As shown in Figure~\ref{fig:pl2d-fq}, in the MAD-0.9 and MAD-0 models, the disk surface is strongly flattened near the black hole, whereas the disk opens up at larger radii.
This indicates that magnetic compression is effective mainly in the inner region.
%We therefore define the MAD radius, $r_\mathrm{MAD}$, using the radius at which the local slope of the disk surface becomes positive.
We therefore define the MAD radius, $r_\mathrm{MAD}$, as the radius where the local slope of the disk surface becomes positive.
Specifically, from the scale height $\theta_H(r)$, we calculate the disk-surface coordinates, $R_H = r\cos\theta_H$ and $z_H = r\sin\theta_H$.
In this procedure, we restrict the analysis to $r>r_\mathrm{hor}$.
At each radius, we perform a linear fit,
$z_H = a R_H + b$,
over the interval
$[R_H-\delta R,\ R_H+\delta R]$.
Here, $\delta R$ corresponds to five grid cells on each side of $R_H$.
Using the local slope $a=a(R)$ obtained from this procedure, we define the MAD transition region as the radial range where $a(R)$ takes values between $0.1$ and $0.5$ times the slope at $R=30r_\mathrm{g}$, $a(30r_\mathrm{g})$. We denote the representative scale of this region by $r_\mathrm{MAD}$. Inside this region, the disk is considered to be in the MAD state. We confirm that the resulting $r_\mathrm{MAD}$ depends only weakly on the number of grid cells used to determine $\delta R$ and the reference radius $R=30r_\mathrm{g}$.
Although these characteristic radii are defined using different operational criteria, their relative ordering provides useful insight into how magnetic fields and radiation regulate the disk structure and outflows.

\begin{figure}
 \begin{center}
\includegraphics[width=7cm]{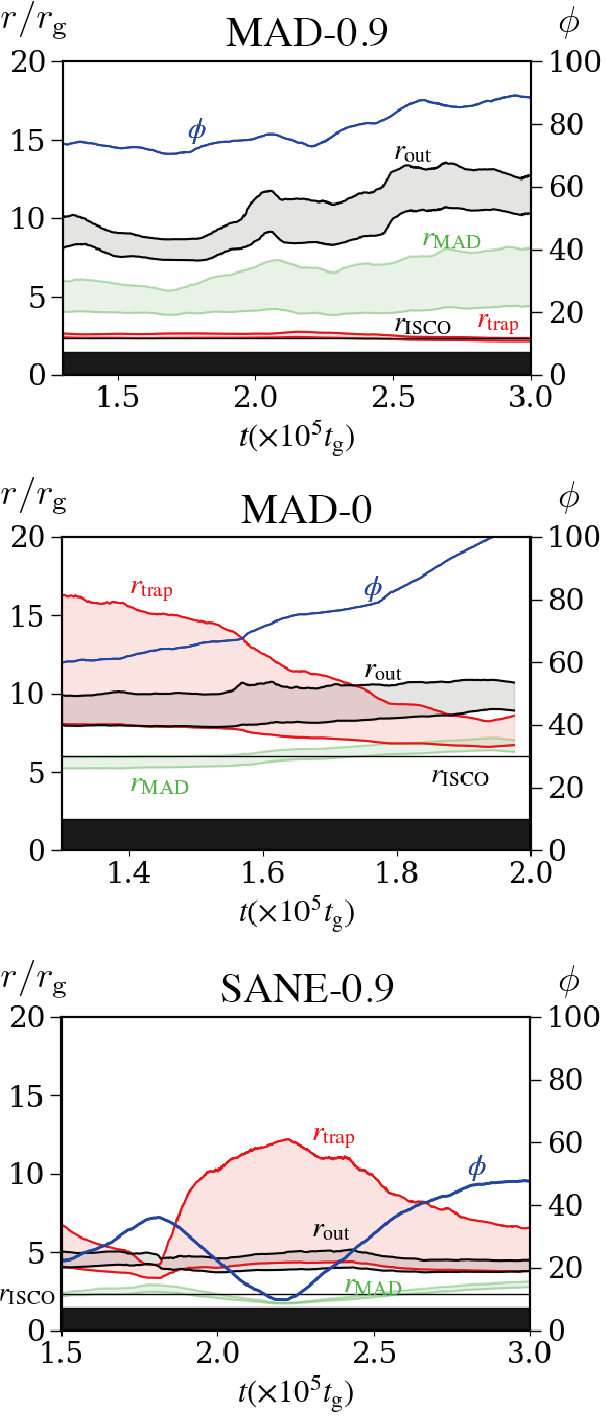}
        \caption{
Time evolution of $r_\mathrm{trap}$ (red), $r_\mathrm{out}$ (black), and $r_\mathrm{MAD}$ (green) for the MAD-0.9 (top), MAD-0 (middle), and SANE-0.9 (bottom) models.
The MAD parameter $\phi$ (blue, right axis) is overplotted to illustrate the correlation between $\phi$ and each characteristic radius, although its time evolution is already shown in Figure~\ref{fig:t-mdot}.
  }
    \label{fig:t-r}
 \end{center}
\end{figure}

%--------------------------------------------%
% 5
%--------------------------------------------%
Figure~\ref{fig:t-r} shows the time evolution of the representative radii, $r_\mathrm{trap}$ (red), $r_\mathrm{out}$ (black), and $r_\mathrm{MAD}$ (green), for the trapping, outflow development, and MAD transition regions, respectively.
From top to bottom, the panels correspond to MAD-0.9, MAD-0, and SANE-0.9, respectively. The black shaded region indicates the region inside the horizon radius.
Although the time evolution of the MAD parameter $\phi$ is already shown in Figure~\ref{fig:t-mdot}, we also overlay it as the blue curve in this figure to clarify the correlation between $\phi$ and the representative radii. All data are smoothed by a moving time average. The averaging window is set to be approximately the viscous time at $r=15r_\mathrm{g}$,
$\sim 4.06\times 10^{4} t_\mathrm{g}
(r/15r_\mathrm{g})^{3/2}
(\alpha/0.1)^{-1}
(\theta_H/0.3)^{-2}$.

%-----------------------------------------------------%
%図8 t-rMAD, r_out, r_trap
%-----------------------------------------------------%

%--------------------------------------------%
% 6
%--------------------------------------------%
In the MAD-0.9 model, a geometrically compressed disk is steadily formed near the black hole, and $r_\mathrm{MAD}$ is located outside the ISCO radius. The value of $r_\mathrm{MAD}$ also shows a positive correlation with $\phi$. This corresponds to the picture in which, as the magnetic flux becomes stronger, the disk is compressed by magnetic pressure and the magnetically dominated region extends to larger radii.
The trapping radius $r_\mathrm{trap}$ shows a negative correlation with $\phi$, 
but remains close to the ISCO radius.
As shown in Section~\ref{result}, electromagnetic energy generated by the Blandford--Znajek mechanism is dissipated near the ISCO, producing additional radiative energy in the inner disk. 
This additional radiative component enhances $L_\mathrm{rad}^\mathrm{out}$ near the black hole, so that the radius where $L_\mathrm{rad}^\mathrm{in}=L_\mathrm{rad}^\mathrm{out}$ is located closer to the ISCO.
The outflow radius $r_\mathrm{out}$ is located slightly outside these radii and shows a positive correlation with $\phi$. This suggests that, in the region of $r<r_\mathrm{MAD}$, where the disk is confined by strong magnetic flux, a substantial mass outflow is less likely to form, whereas a disk wind develops at $r>r_\mathrm{MAD}$.

%--------------------------------------------%
% 7
%--------------------------------------------%
In the MAD-0 model, the value of $r_\mathrm{MAD}$ is comparable to that in MAD-0.9. However, because the black hole spin is $a_*=0$, $r_\mathrm{MAD}$ is located near the ISCO radius. The trapping radius $r_\mathrm{trap}$ is located outside the ISCO radius and is larger than that in MAD-0.9. In MAD-0.9, $r_\mathrm{trap}$ becomes smaller because of disk heating by the Blandford--Znajek mechanism. In MAD-0, in contrast, this heating mechanism does not operate, and the representative radius of the trapping region is therefore larger than that in MAD-0.9.
The trapping radius $r_\mathrm{trap}$ also shows a negative correlation with $\phi$. This may correspond to the fact that the disk becomes thinner even at $r\gtrsim r_\mathrm{MAD}$ owing to the effect of magnetic flux, which shortens the photon diffusion time from the disk interior to the outside \citep{McKinneyEfficiencySuperEddington2015}. As in MAD-0.9, $r_\mathrm{out}$ is located outside $r_\mathrm{MAD}$ and has a radius comparable to $r_\mathrm{trap}$. This suggests that the disk wind is mainly driven by radiation \citep{TakeuchiNovelJetModel2010,TakahashiRadiationDragEffects2015,MollerAccelerationWindOptically2015}.

%--------------------------------------------%
% 8
%--------------------------------------------%
In the SANE-0.9 model, $r_\mathrm{MAD}$ is located close to the ISCO radius.
This should not be interpreted as evidence for a well-developed MAD region extending down to the ISCO, because $r_\mathrm{MAD}$ is defined operationally from the disk-surface geometry rather than directly from the local magnetic flux or magnetization.
Thus, the proximity of $r_\mathrm{MAD}$ to the ISCO in SANE-0.9 should be regarded as a geometrical indicator.
In this model, the time variability of $\phi$ is large because more magnetic loops are initially placed than in MAD-0.9 and MAD-0. Nevertheless, $r_\mathrm{trap}$ shows a negative correlation with $\phi$.
At $t\simeq 2.2\times10^{5} t_\mathrm{g}$, $\phi$ is small, and correspondingly the trapping radius is large. As $\phi$ subsequently increases, $r_\mathrm{trap}$ becomes smaller. As in MAD-0, this likely reflects the more efficient cooling caused by disk thinning. 
In addition, as $\phi$ increases, electromagnetic energy dissipation driven by the Blandford--Znajek mechanism may begin to operate effectively.
The outflow radius $r_\mathrm{out}$ is comparable to $r_\mathrm{trap}$, as in MAD-0, suggesting that the outflow may be driven mainly by radiation.

%--------------------------------------------%
% 9
%--------------------------------------------%
To further illustrate the role of the magnetic flux parameter $\phi$ as an organizing variable,
we replot the characteristic radii as functions of $\phi$ in Appendix \ref{sec:phi-r}.

%********************************************%
% section 4.2
%********************************************%
\subsection{From trapping radius to observables}
%--------------------------------------------%
% 1
%--------------------------------------------%

%--------------------------------------------%
% 1
%--------------------------------------------%
In this study, we find that the trapping radius depends on both the magnetic flux threading the disk and the spin parameter $a_*$. As shown in Figure~\ref{fig:pl2d-fq}, the trapping radius is located inside the photosphere (yellow line) and is therefore not a directly observable quantity. Nevertheless, our results suggest that the trapping radius plays an important role in determining the radiative properties of supercritical accretion disks.

%--------------------------------------------%
% 2
%--------------------------------------------%
We mainly focus on the radiative luminosity and energy budget, and show that the magnetic flux is negatively correlated with the trapping radius. As the trapping radius becomes smaller with increasing magnetic flux, the relative contribution of radiation from hotter regions may increase, potentially leading to spectral hardening.
Even for the same amount of magnetic flux, a higher black hole spin reduces the ISCO radius and brings it closer to the horizon. The contribution of radiation from the inner disk region is therefore expected to increase, resulting in a harder spectrum.

%--------------------------------------------%
% 3
%--------------------------------------------%
Furthermore, the rotational energy of the black hole extracted by the Blandford--Znajek mechanism is dissipated near the ISCO and converted into radiative energy. As shown in Figure~\ref{fig:pl2d-fq}, the effective optical depth near the black hole is below unity. Under such conditions, the dissipated energy is expected to contribute efficiently to the radiation field through Comptonization. In particular, for high-spin black holes with large magnetic flux, magnetic power comparable to the accretion power can be generated \citep{TchekhovskoyEfficientGenerationJets2011}, potentially producing a hard spectrum \citep{TakahashiFORMATIONOVERHEATEDREGIONS2016}.

%--------------------------------------------%
% 4
%--------------------------------------------%
These effects may be related to the observed properties of systems showing supercritical accretion, particularly ultraluminous X-ray sources (ULXs). These objects are known to exhibit diverse observed spectra \citep{KaaretUltraluminousRaySources2016}.
%, even though similar Eddington ratios are suggested . 
Our results suggest that differences in the magnetic-flux state of the disk may be one of the factors producing such diversity.

%--------------------------------------------%
% 5
%--------------------------------------------%
On the other hand, both high black hole spin and large magnetic flux are expected to harden the spectrum. Therefore, a degeneracy can arise when one attempts to estimate the magnetic flux and black hole spin simultaneously from the observed spectral properties. However, this degeneracy may be alleviated by focusing on time-variable radiative properties.
Black hole spin evolves only on long timescales through the accumulation of mass and angular momentum by accretion \citep{BardeenKerrMetricBlackHoles1970,ThorneDiskAccretionOnto1974,GammieBlackHoleSpin2004,VolonteriBlackHoleSpin2007}.
In contrast, the magnetic-flux state can vary with time, depending on the growth of magnetorotational instability, the transport and release of magnetic flux, and the supply of magnetic fields from the external environment. Therefore, by interpreting the observed properties together with time-variable radiative features and state transitions, it may be possible to observationally distinguish whether the disk is close to the MAD state.

%--------------------------------------------%
% 6
%--------------------------------------------%
Finally, the observed spectrum is also affected by the viewing angle, absorption and scattering by disk winds, and the electron temperature and nonthermal processes. Detailed spectral formation calculations are therefore required to evaluate these effects quantitatively. Our results suggest that the magnetic-flux state can play an important role in spectral formation, and a detailed investigation of this issue is left for future work.

%********************************************%
% section 5
%********************************************%
\section{Summary}\label{summary}
%--------------------------------------------%
% 1
%--------------------------------------------%
In this study, we investigate how the amount of magnetic flux and black hole spin affect energy transport, dissipation structures, and radiative properties in supercritical accretion flows using general relativistic radiation magnetohydrodynamic simulations initialized with a torus placed far from the black hole. In particular, we focus primarily on the high-spin MAD state with strong magnetic flux. By comparing this model with a non-spinning MAD state and a high-spin SANE state, we disentangle the contributions of magnetic flux and black hole spin.

%--------------------------------------------%
% 2
%--------------------------------------------%
Our analysis shows that, in the high-spin MAD state with strong magnetic flux, a strong Poynting flux supplied from the black hole is transported along the region near the disk surface in the vicinity of the black hole. A fraction of the transported electromagnetic energy is dissipated in this region and contributes significantly to the generation of radiative energy. In addition, a magnetized outflow is formed from this region, and the generated radiative energy is transported outward together with the outflow. As a result, the time-averaged radiative efficiency reaches $\eta_\mathrm{rad}=0.60$.
In contrast, in the high-spin SANE state, the electromagnetic energy supply from the black hole is much weaker, and the radiative efficiency remains at $\eta_\mathrm{rad}=0.21$, which is smaller than that in the MAD state. In the non-spinning MAD state, because energy supply by the Blandford--Znajek mechanism does not occur, the radiative efficiency is $\eta_\mathrm{rad}=0.088$. These comparisons show that the high radiative efficiency of supercritical MAD accretion is not determined solely by the strong magnetic-flux state, but is greatly enhanced by electromagnetic energy supply associated with black hole spin.

%--------------------------------------------%
% 3
%--------------------------------------------%
In the high-spin MAD state, the additional outward radiative component results in a small effective trapping radius of $r_\mathrm{trap}=2.5r_\mathrm{g}$, comparable to the ISCO radius. In contrast, in the high-spin SANE state, electromagnetic energy transport originating from the Blandford--Znajek mechanism and the associated outward radiative component are weaker, and the effective trapping radius is larger than that in the MAD state.
We also find that the trapping radius shows a negative correlation with the magnetic flux in both the high-spin MAD and high-spin SANE states, and a similar trend is also found in the non-spinning MAD state. Since photon trapping is characterized by a comparison between the photon diffusion time and the inflow time \citep{SadowskiGlobalSimulationsAxisymmetric2015}, this trend suggests that changes in the disk structure and radiation transport associated with magnetic-flux accumulation, as discussed by \citet{McKinneyEfficiencySuperEddington2015}, may affect the effective trapping radius.

%--------------------------------------------%
% 4
%--------------------------------------------%
These results suggest that the radiative properties of supercritical accreting systems, such as ultraluminous X-ray sources, may depend not only on the mass accretion rate, but also on the magnetic flux accumulated on the black hole and the black hole spin.
The mass accretion rate affects spectra through Compton upscattering and downscattering in dense outflows \citep{KawashimaNewSpectralState2009,KawashimaComptonizedPhotonSpectra2012}.
In addition, a rapidly spinning black hole has a smaller ISCO radius, which can enhance the contribution of radiation from the vicinity of the black hole.
When strong magnetic flux is accumulated around such a black hole, electromagnetic energy dissipation near the disk surface can further enhance this inner radiation component.
These effects may produce a harder spectral component.
Because black hole spin evolves only on long timescales, whereas the accumulated magnetic flux can vary with the dynamical and viscous evolution of the accretion flow, time-dependent spectra and luminosities may help distinguish these effects and constrain both the black hole spin and magnetic flux. 
To test this interpretation quantitatively, global three-dimensional GR-RMHD simulations with systematically varied magnetic flux, black hole spin, and mass accretion rate are required.
Detailed spectral calculations based on these simulations are also necessary.

\begin{acknowledgments}
This work was supported by JSPS KAKENHI Grant Numbers JP24K00672 (H.R.T.),
JP24K00678 and JP21H04488 (H.R.T. and K.O.),
JP24KF0130 and JP26KF0011 (K.O.), 
JP25K01045 (K.O. and Y.A.), 
and JP24KJ0143 and JP25K17439 (A.I.).
This work was also supported by MEXT as “Program for Promoting Researches on the Supercomputer Fugaku” (Structure and Evolution of the Universe Unraveled by Fusion of Simulation and AI; Grant Number JPMXP1020240219; Y.A., H.R.T., and K.O.) and by the Joint Institute for Computational Fundamental Science (JICFuS; K.O.).
This research used the computational resources of Pegasus provided by the Multidisciplinary Cooperative Research Program at the Center for Computational Sciences, University of Tsukuba. This research was also conducted using the Supermicro ARS-111GL-DNHR-LCC and FUJITSU Server PRIMERGY CX2550 M7 (Miyabi) at the Joint Center for Advanced High Performance Computing (JCAHPC).
The authors used ChatGPT \citep{OpenAIChatGPT} to assist with English-language editing.
% This work was supported by JSPS KAKENHI Grant Numbers JP24K00672, JP21H04488, JP24K00678 (H.R.T.), JP24KF0130 (K.O.), JP26KF0011 (K.O.), and JP25K01045 (K.O. and Y.A.).
% This work was also supported by MEXT as “Program for Promoting Researches on the Supercomputer Fugaku” (Structure and Evolution of the Universe Unraveled by Fusion of Simulation and AI; Grant Number JPMXP1020240219; Y.A., H.R.T., K.O.), by Joint Institute for Computational Fundamental Science (JICFuS, K.O.).
% This research used computational resources of Pegasus provided by Multidisciplinary Cooperative Research Program 
% in Center for Computational Sciences, University of Tsukuba.
% This research was also conducted using the Supermicro ARS-111GL-DNHR-LCC and FUJITSU Server PRIMERGY CX2550 M7 (Miyabi) 
% at Joint Center for Advanced High Performance Computing (JCAHPC).
% %This work was also supported by MEXT as “Program for Promoting Researches on the Supercomputer Fugaku” (Toward a unified view of the universe: from large-scale structures to planets, JPMXP1020200109; H.R.T., and K. O.) and by Joint Institute for Computational Fundamental Science (JICFuS; K.O.).  
\end{acknowledgments}

\appendix

\section{Definition of effective viscous heating rates 
and effective emission rates}\label{qvis}
%------------------------------
%------------------------------
In this appendix, we describe the definitions of the effective viscous heating rate $q_{\mathrm{vis}}$ and the effective emission rate $q_{\mathrm{emi}}$ used in this paper.
We solve the general relativistic radiation magnetohydrodynamic (GR-RMHD) equations using a semi-implicit method \citep{TakahashiExplicitImplicitScheme2013,TakahashiNumericalTreatmentAnisotropic2013}.
The basic equations can be written symbolically as
\begin{equation}
\frac{\partial U_a}{\partial t} + \nabla_j F_a^j = S_a,
\label{ap:eq}
\end{equation}
where $U_a$, $F_a^j$, and $S_a$ denote the conserved variables, fluxes, and source terms, respectively.
Here, the index $a$ labels each evolved equation, including the mass, energy-momentum, induction, radiation energy-momentum, and entropy equations.
The entropy equation is used as a fallback when the conversion from conserved variables to primitive variables becomes difficult \citep{McKinneyThreeDimensionalGeneral2014}.
In this study, we also use this entropy conservation equation to evaluate the effective viscous heating rate.

%------------------------------
%------------------------------
We solve Equation~(\ref{ap:eq}) using operator splitting as follows:
\begin{eqnarray}
&&\frac{\partial U_a}{\partial t} + \nabla_j F_a^j = 0,
\label{ap:eq2}\\
&&\frac{\partial U_a}{\partial t} = S_a.
\label{ap:eq3}
\end{eqnarray}
Equation~(\ref{ap:eq2}) is integrated explicitly in time, and Equation~(\ref{ap:eq3}) is then solved using an implicit method.
Equation~(\ref{ap:eq2}) describes the source-free evolution, including advection and compression of gas, magnetic fields, and radiation.
The change in the gas energy obtained in this step also includes heating due to numerically captured viscous and magnetic dissipation.
Equation~(\ref{ap:eq3}) describes the local interaction between gas and radiation through the radiation four-force.

%------------------------------
%------------------------------
Let $e^n$ be the gas energy at time step $n$.
We denote the gas energy obtained by solving the energy conservation equation in Equation~(\ref{ap:eq2}) by $e^{\mathrm{ene}}$, and the gas energy obtained from the entropy conservation equation in the same step by $e^{\mathrm{ent}}$.
We also denote the gas energy at time step $n+1$, after solving Equation~(\ref{ap:eq3}), by $e^{n+1}$.

%------------------------------
%------------------------------
We then define the effective emission rate as
\begin{equation}
q_{\mathrm{emi}} =
\frac{e^{\mathrm{ene}} - e^{n+1}}{\Delta t}.
\end{equation}
Here, $\Delta t$ is the time interval from step $n$ to step $n+1$.
This quantity represents the energy transferred from gas to radiation per unit time through radiative processes.
We note that $q_{\mathrm{emi}}$ is a local energy-transfer rate, and that the generated radiation does not necessarily escape from the disk.
Therefore, its volume integral does not necessarily coincide with the effective radiative cooling rate used in conventional disk models.

%------------------------------
%------------------------------
Next, we define the effective viscous heating rate.
The difference $e^{\mathrm{ene}} - e^n$ includes heating due to viscous and magnetic dissipation, in addition to advection and compression.
On the other hand, because the entropy conservation equation assumes an adiabatic process, the difference $e^{\mathrm{ent}} - e^n$ does not include these dissipative heating contributions.
We therefore define the effective viscous heating rate in this paper as
\begin{equation}
q_{\mathrm{vis}} =
\frac{e^{\mathrm{ene}} - e^{\mathrm{ent}}}{\Delta t}.
\end{equation}
Similarly, the contribution from advection and compression is evaluated as
\begin{equation}
q_{\mathrm{adv/comp}} =
\frac{e^{\mathrm{ent}} - e^n}{\Delta t}.
\end{equation}

%------------------------------
%------------------------------
Finally, we describe the definition of the gas energy $e$ used in this paper.
The simplest choice is to use the internal energy $e_{\mathrm{int}}$, which is a natural definition when discussing the thermal balance of the disk.
Alternatively, we define the gas energy using the gas contribution to the energy-momentum tensor:
\begin{equation}
e =
- \left[
(p_{\mathrm{gas}} + e_{\mathrm{int}})\,u_t u^t
+ p_{\mathrm{gas}}
\right]     
  \label{ap:tensor}
\end{equation}
The advantage of this definition is that it is derived consistently from the energy-momentum conservation law.

\begin{figure}
 \begin{center}
\includegraphics[width=8cm]{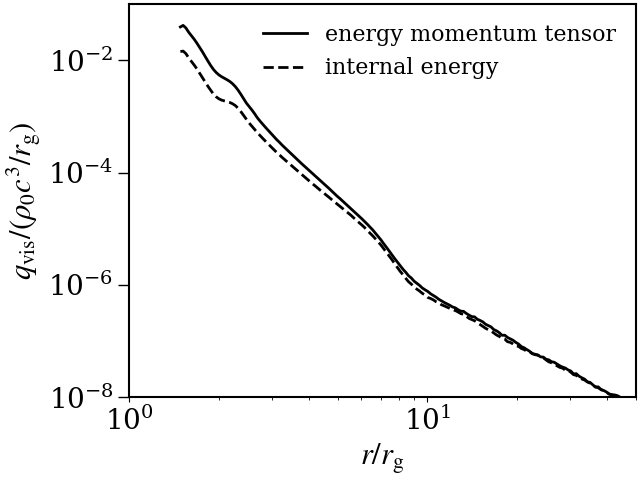}
        \caption{
Radial profiles of the effective viscous heating rate, normalized by $\rho_0 c^3 / r_\mathrm{g}$.
Solid curve shows the heating rate evaluated from the energy-momentum tensor (see Equation~\ref{ap:tensor}), while dashed curve shows that calculated from the internal energy density $e_\mathrm{int}$.
}
    \label{fig:r-qvis-test}
 \end{center}
\end{figure}

%------------------------------
%------------------------------
Figure~\ref{fig:r-qvis-test} shows the radial profiles of the effective viscous heating rate in the MAD-0.9 model.
The profiles are time-averaged over $t=100{,}000\text{--}130{,}000\,t_\mathrm{g}$ and averaged over the $\theta$ and $\phi$ directions with density weighting.
Solid lines show $q_{\mathrm{vis}}$ evaluated based on the energy-momentum tensor, whereas dashed lines show $q_{\mathrm{vis}}$ evaluated based on the internal energy $e_{\mathrm{int}}$.
Both the vertical and horizontal axes are shown in logarithmic scale.

%------------------------------
%------------------------------
Near the black hole, especially in the region of $r \lesssim 3\text{--}4\,r_{\mathrm g}$, $q_{\mathrm{vis}}$ evaluated from the energy-momentum tensor becomes larger than that evaluated from the internal energy by up to a factor of about 2.9.
On the other hand, sufficiently far from the black hole, there is little difference between the two estimates, and the radial profiles of the viscous heating rate are almost identical.

%------------------------------
%------------------------------
The purpose of this study is to clarify the origin of the radiation generated by the dissipation of Poynting flux.
From this perspective, $q_{\mathrm{vis}}$ defined based on energy conservation is considered to be a more physically appropriate diagnostic.
Therefore, in this paper, we adopt $q_{\mathrm{vis}}$ evaluated from the energy-momentum tensor.
%-------------------------------------%
% Appendix A Figure 
%-------------------------------------%

%******************************
% Appendix B. rのphi依存性
%******************************
\section{MAD parameter dependence of characteristic radii}\label{sec:phi-r}
%------------------------------
%------------------------------
In section \ref{sec:t-r}, we show the time evolution of the characteristic radii, $r_\mathrm{trap}$, $r_\mathrm{out}$, and $r_\mathrm{MAD}$, and demonstrate that these radii vary in correlation with the magnetic flux $\phi$.
In this appendix, we reorganize the same results by using $\phi$ as the horizontal axis, in order to visualize the relation between the characteristic radii and $\phi$ in a form that does not explicitly depend on time evolution.
This figure is intended to provide a supplementary view of the trends discussed in the main text.
% This figure does not present new physical results, but is intended to provide a supplementary confirmation of the trends discussed in the main text.
Additional models with a smaller initial torus size are also shown as a consistency check for these trends.

\begin{figure}
 \begin{center}
\includegraphics[width=14cm]{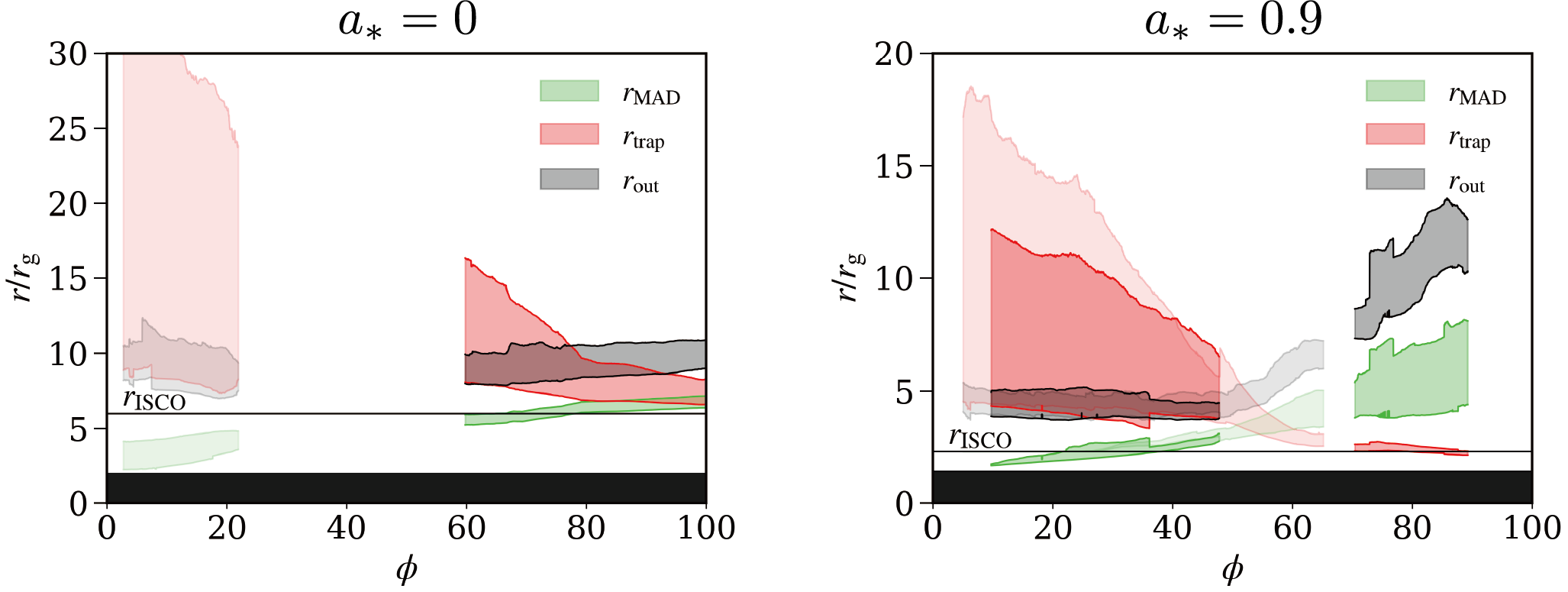}
        \caption{
  Three characteristic radii, $r_{\rm trap}$ (red), $r_{\rm out}$ (black),
  and $r_{\rm MAD}$ (green) are plotted as functions of $\phi$.
  Left and right panels correspond to $a_*=0$ and $a_*=0.9$, respectively.
  The horizontal black line indicates the ISCO radius.
  Shaded regions represent the envelopes of the characteristic radii, 
  reflecting the scatter caused by time variability and fluctuations in the simulations at similar values of $\phi$.
  Results obtained with modified initial conditions (small-0 and small-0.9)
  are shown in lighter colors to distinguish them from the main models.         
}
    \label{fig:phi-r}
 \end{center}
\end{figure}

%------------------------------
%------------------------------
Figure~\ref{fig:phi-r} shows the dependence of $r_\mathrm{trap}$ (red), $r_\mathrm{out}$ (black), and $r_\mathrm{MAD}$ (green) on $\phi$.
The left and right panels show the results for $a_*=0$ and $a_*=0.9$, respectively.
The horizontal black line indicates the ISCO radius.
Because of time variability and fluctuations in the simulations, each characteristic radius does not take a single value even for a similar MAD parameter, but instead has a finite spread.
To reflect this scatter, we show the radii using envelopes that cover the maximum radial range allowed at each $\phi$.
In this figure, models with the same black hole spin are plotted in the same panel.
In particular, MAD-0.9 and SANE-0.9 are shown together in the right panel as models with $a_*=0.9$, while the models with $a_*=0$ are shown together in the left panel.
This allows us to compare the behavior of the characteristic radii across models using the magnetic flux $\phi$, rather than time, as the organizing variable.

%------------------------------
%------------------------------
In addition to the main models used in this paper, we also show models with modified initial conditions as a supplementary check of the robustness of the main results.
Specifically, in the main models, the inner edge of the initial torus and the radius of the pressure maximum are set to $r_{\rm edge}=200r_{\rm g}$ and $r_{\rm peak}=300r_{\rm g}$, respectively, whereas in the models with modified initial conditions, they are set to $r_{\rm edge}=20r_{\rm g}$ and $r_{\rm peak}=33r_{\rm g}$, respectively.
The number of initial magnetic loops is set to $N_{\rm mag}=1$.
Among the models with these initial conditions, we refer to the cases with black hole spins of $a_*=0$ and $a_*=0.9$ as small-0 and small-0.9, respectively.
The mass accretion rates averaged over $t=5000\,t_\mathrm{g}$--$300{,}000\,t_\mathrm{g}$ are $270\, L_\mathrm{Edd}$ and $530\, L_\mathrm{Edd}$, respectively.
In these models, the mass accretion rate shows a slight decreasing trend with time, likely because of the smaller initial torus size.
However, this trend is not expected to significantly affect the relation between the characteristic radii and the magnetic flux $\phi$ discussed in this paper.
Indeed, small-0 shows a SANE state, while small-0.9 evolves from a SANE state to a state close to MAD, and these models therefore complement the main model set used in this paper.
The $\phi$ dependence of the characteristic radii obtained from these models is also broadly consistent with the trends found in the main models.
These results are shown in lighter colors in Figure~\ref{fig:phi-r} to distinguish them from the main models.

%------------------------------
%------------------------------
First, we consider the case of $a_*=0$.
In the region of low magnetic flux $\phi$, $r_\mathrm{trap}$ takes relatively large values, and the trapping transition region extends over a wide radial range.
This can be understood as a consequence of defining the trapping transition region by $L_\mathrm{rad}^\mathrm{out}/L_\mathrm{rad}^\mathrm{in}=0.5$--1.
As $r_\mathrm{trap}$ approaches the ISCO, the gravitational potential becomes rapidly deeper, and the energy released per unit radius also increases.
As a result, $L_\mathrm{rad}^\mathrm{out}$ increases rapidly with decreasing radius, and the radial separation between the locations corresponding to $L_\mathrm{rad}^\mathrm{out}=0.5\,L_\mathrm{rad}^\mathrm{in}$ and $L_\mathrm{rad}^\mathrm{out}=L_\mathrm{rad}^\mathrm{in}$ becomes small.
On the other hand, when $r_\mathrm{trap}$ is located farther away from the ISCO, the gravitational potential changes more gradually, and the radial dependence of $L_\mathrm{rad}^\mathrm{out}$ is also weaker.
The trapping transition region therefore corresponds to a wider radial range.

%------------------------------
%------------------------------
As shown in Section~\ref{sec:t-r}, $r_\mathrm{trap}$ systematically becomes smaller as the magnetic flux $\phi$ increases.
The MAD radius $r_\mathrm{MAD}$ also increases with $\phi$, and becomes comparable to the ISCO radius in the MAD regime.
On the other hand, $r_\mathrm{out}$ shows a relatively weak dependence on $\phi$.
This may indicate that the formation of the outflow is driven mainly by radiation pressure \citep{TakeuchiNovelJetModel2010,TakahashiRadiationDragEffects2015,MollerAccelerationWindOptically2015}.

%------------------------------
%------------------------------
Next, we consider the case of $a_*=0.9$.
As in the case of $a_*=0$, $r_\mathrm{trap}$ systematically becomes smaller as the magnetic flux $\phi$ increases.
The MAD radius $r_\mathrm{MAD}$ also increases with $\phi$, and a clear MAD region is formed in the high-$\phi$ regime.
This behavior suggests that, in the high-spin case, an increase in magnetic flux may extend the radial range that contributes to the outflow.

%------------------------------
%------------------------------
These results indicate that the $\phi$ dependence of $r_\mathrm{trap}$ and $r_\mathrm{MAD}$ is qualitatively similar between $a_*=0$ and $a_*=0.9$, whereas $r_\mathrm{out}$ may show spin-dependent behavior.

% \section{Using Chinese, Japanese, and Korean characters}
% Authors have the option to include names in Chinese, Japanese, or Korean (CJK)
% characters in addition to the English name. The names will be displayed
% in parentheses after the English name. The way to do this in AASTeX is to
% use the CJK package available at \url{https://ctan.org/pkg/cjk?lang=en}.
% Further details on how to implement this and solutions for common problems,
% please go to \url{https://journals.aas.org/nonroman/}.

%% For this sample we use BibTeX plus aasjournalv7.bst to generate the
\bibliographystyle{aasjournalv7.1}
\bibliography{hrtakahashi}



\end{document}